\documentclass[12pt,draftcls, peerreview, onecolumn]{IEEEtran}
\usepackage{cite}
\usepackage{graphicx}
\usepackage{url}
\usepackage[centertags]{amsmath}
\usepackage{algorithm}
\usepackage{algorithmic}
\usepackage{amssymb}
\usepackage{epsfig,color}
\newcommand{\beq}{\begin{equation}}
\newcommand{\eeq}{\end{equation}}
\newcommand{\beqn}{\begin{align}}
\newcommand{\eeqn}{\end{align}}

\usepackage{mathtools}

\DeclarePairedDelimiter{\floor}{\lfloor}{\rfloor}
\begin{document}

\title{Analysis and Design of  Secure Massive MIMO Systems in the Presence of Hardware Impairments}

\author{\IEEEauthorblockN{Jun~Zhu, Derrick~Wing~Kwan~Ng, Ning~Wang, Robert~Schober, and Vijay~K.~Bhargava,}\\
\IEEEauthorblockA{The University of British Columbia\\
}

\thanks{%
This work has been in part presented at The 17th IEEE International Workshop on Signal Processing Advances in Wireless Communications 2016 (SPAWC 2016) \cite{SPAWC}.}
}
%

\IEEEoverridecommandlockouts

\setcounter{page}{1}

\maketitle

\begin{abstract}
To keep the hardware costs of future communications systems manageable, the use of low-cost hardware components is desirable. This is particularly true for the emerging massive multiple-input multiple-output (MIMO)
systems which equip base stations (BSs) with a large number of antenna elements. However, low-cost transceiver designs will further accentuate the hardware impairments which are present in any practical communication
system. In this paper, we investigate the impact of hardware impairments on the secrecy performance of downlink massive MIMO systems in the presence of a passive multiple-antenna eavesdropper. Thereby,
for the BS and the legitimate users, the joint effects of multiplicative phase noise, additive distortion noise, and amplified receiver noise are taken into account, whereas the eavesdropper is assumed to employ
ideal hardware. We derive a lower bound for the ergodic secrecy rate of a given user when matched filter (MF) data precoding and artificial noise (AN) transmission are employed at the BS. Based on the derived
analytical expression, we investigate the impact of the various system parameters on the secrecy rate and optimize both the pilot sets used for uplink training and the AN precoding.
Our analytical and simulation results reveal that 1) the additive distortion noise at the BS may be beneficial for the secrecy performance, especially if the power assigned for AN emission is not sufficient;
 2) all other hardware impairments have a negative impact on the secrecy performance; 3) {{despite their susceptibility to pilot interference in the presence of phase noise}}, so-called spatially orthogonal pilot sequences are preferable unless the phase noise is very strong; 4) the proposed
 generalized null-space (NS) AN precoding method can efficiently mitigate the negative effects of phase noise.
\end{abstract}

\section{Introduction}
The emerging massive multiple-input multiple-output (MIMO) architecture promises tremendous performance gains in terms of network throughput and energy efficiency
by employing simple coherent processing across arrays of hundreds or even thousands of base station (BS) antennas, serving tens or hundreds of mobile terminals \cite{survey,noncooperative}. Thereby, physical layer channel impairments such as fading, additive Gaussian noise, and interference are averaged out in the limit of an infinite number of antennas \cite{survey}-\cite{pilotcontam}. As an additional benefit, massive MIMO is inherently more
secure than conventional MIMO systems, as the large-scale antenna array equipped at the transmitter (Alice) can accurately focus a narrow and directional information beam on the intended terminal (Bob), such that the received signal power at Bob is several orders of magnitude higher than that at any incoherent passive eavesdropper (Eve) \cite{phymassive}. Unfortunately, this benefit may vanish if Eve also employs a massive antenna array for eavesdropping. In this case, unless additional measures to secure the communication are taken by Alice, even a single passive Eve may be able to intercept the signal intended for Bob \cite{zhu,zhu_letter}.

Since security is a critical concern for future communication systems, facilitating secrecy at the physical layer of (massive) MIMO systems has received significant
attention recently. {{Physical layer security for conventional (non-massive) MIMO transmission has been extensively studied in the literature, e.g. \cite{R1,R2,R3}.}} A large system secrecy analysis of MIMO systems was provided in \cite{massivesec,largerci}. Thereby, availability of Eve's channel state
information (CSI) at Alice was assumed, which is an optimistic assumption in practice. Artificial noise (AN) generation \cite{negi} was employed to provide physical
layer security in a multi-cell massive MIMO system with pilot contamination in \cite{zhu} for the case when Eve's CSI is not known. Thereby, it was shown that secure
communication can be achieved even with simple matched filter (MF) precoding of the data and null-space (NS) precoding of the AN.  Nevertheless, it was revealed in \cite{zhu2}
that significant additional performance gains are possible with more sophisticated data and AN precoders, including polynomial precoders. Furthermore, AN-aided jamming of Rician
fading massive MIMO channels was investigated in \cite{jwang}. In the context of massive MIMO relaying, the work presented in \cite{chen1} compared two classic relaying schemes, i.e., amplify-and forward (AF) and decode-and-forward (DF), for physical layer security with imperfect CSI at the massive MIMO relay. While \cite{zhu}-\cite{chen1} assumed that Eve is passive, the so-called pilot contamination attack, a form of active
eavesdropping, was also considered in the literature. In particular,  several techniques for detection of  the pilot contamination attack were proposed in \cite{phymassive}. Moreover, the authors in \cite{SKA}
developed a secret key agreement protocol under the pilot contamination attack, and the authors in \cite{phymassive2} proposed to encrypt the pilot sequence in order to
hide it from the attacker. Several techniques for combating the pilot contamination attack at the physical layer of a multi-cell massive MIMO system were proposed in \cite{wyp}.

All aforementioned works on secure massive MIMO are based on the assumption that the transceivers of the legitimate users are equipped with perfect hardware components, i.e.,
the effects of hardware impairments (HWIs) were not taken into account. Nevertheless, all practical implementations do suffer from HWIs such as phase noise, quantization
errors, amplification noise, and nonlinearities \cite{nonideal}. These impairments are expected to be particularly pronounced in massive MIMO systems as the excessive number of
BS antennas makes the use of low-cost components desirable to keep the overall capital expenditures for operators manageable. Although HWIs can be mitigated
by analog and digital signal processing techniques \cite{aaa}, they cannot be removed completely, due to the randomness introduced by the different sources of imperfection. The
remaining residual HWIs can be modelled by a combination of phase noise and additive distortion noises at the transmitter and the receiver \cite{aaa}. Several works
have investigated the impact of HWIs on massive MIMO systems \cite{nonideal}, \cite{pn}-\cite{MRC}. The impact of phase noise originating from free-running oscillators
on the downlink performance of massive MIMO systems was studied in \cite{pn} for different linear precoder designs. Constant envelope precoding for massive MIMO was studied in \cite{CE} and \cite{zhu_CE}
with the objective of avoiding distortions caused by power amplifier nonlinearities at the transmitter. The impact of the aggregate effects of several HWIs originating from
different sources on massive MIMO systems was studied in \cite{nonideal} by modelling the residual impairments remaining after compensation as additive distortion noises \cite{aaa}.
The authors in \cite{MRC} presented closed-form expressions for the achievable user rates in uplink massive MIMO systems for a general residual HWI model including
both multiplicative phase noise and additive distortion noise. The aforementioned works demonstrated that HWIs can severely limit the performance of massive MIMO
systems. Thereby, a crucial role is played by the degradation caused by phase noise to the quality of the CSI estimates needed for precoder design. On the one hand, phase noise
causes the CSI estimates to become outdated more quickly. On the other hand, it may cause a loss of orthogonality of the pilot sequences employed by the different users
in a cell for uplink training. To overcome the latter effect, so-called temporally orthogonal (TO) and spatially orthogonal (SO) pilot sequences were investigated in  \cite{MRC}.
Furthermore, the impact of the number of local oscillators (LOs) employed at the massive MIMO BS on performance in the presence of phase noise was studied in \cite{pn,MRC}, {while the effect of HWIs on full-duplex massive MIMO relaying was considered in \cite{R4}.}

Communication secrecy aspects are not adequately considered in existing works studying the impact of HWIs in the context of massive MIMO system design \cite{nonideal}, \cite{pn}-\cite{R4}. However, if communication secrecy is considered, an additional challenge arises: Whereas the legitimate user of the system will likely employ
low-cost equipment giving rise to HWIs, the eavesdropper is expected to employ high-quality HWI-free equipment. This disparity in equipment quality was not considered in the related work on physical layer security \cite{zhu}-\cite{wyp}
nor in the related work on HWIs \cite{nonideal}, \cite{pn}-\cite{MRC} and necessitates the development of a new analysis and design framework. For example,
NS AN precoding, which was widely used to enhance the achievable secrecy rate of massive MIMO systems \cite{zhu,zhu2,zhou}, becomes ineffective in the presence of
phase noise.

Motivated by the above considerations, in this paper, we present the first study of physical layer security in hardware constrained massive MIMO systems. Thereby, we focus on the
downlink and adopt for the legitimate links the generic residual HWI model from \cite{aaa,MRC}, which includes the effects of multiplicative phase noise and additive
distortion noise at the BS and the users. As a worst-case scenario, the eavesdropper is assumed to employ ideal hardware. Our main contributions are summarized as follows.
\begin{itemize}
\item For the adopted generic residual HWI model, we derive a tight lower bound for the ergodic secrecy rate achieved by a downlink user when MF data precoding
is employed at the massive MIMO BS. The derived bound provides insight into the impact of various system and channel parameters, such as the phase noise variance,
the additive distortion noise parameters, the AN precoder design, the amount of power allocated to the AN, the pilot sequence design, the number of deployed LOs, and the number of users,
on the ergodic secrecy rate.
\item As conventional NS AN precoding is sensitive to phase noise, we propose a novel generalized NS (G-NS) AN precoding design, which mitigates the AN leakage caused to
the legitimate user in the presence of phase noise at the expense of a reduction of the available spatial degrees of freedom. The proposed method leads to significant performance
gains, especially in systems with large numbers of antennas at the BS.
\item We generalize the SO and TO pilot sequence designs from \cite{MRC} to orthogonal pilot sequences with arbitrary numbers of non-zero elements. Although SO sequences,
which have no zero elements, are preferable for small phase noise variances, sequence designs with zero elements become beneficial in the presence of strong phase noise.
\item Our analytical and numerical results reveal that while HWIs in general degrade the achievable secrecy rate, the proposed countermeasures are effective in limiting this
degradation. Furthermore, surprisingly, there are cases when the additive distortion noise at the BS is beneficial for the secrecy performance as it can have a similar effect as AN.
\end{itemize}

The remainder of this paper is organized as follows. In Section \ref{s2}, the models for uplink training and downlink data transmission in the considered massive MIMO system with
imperfect hardware are presented. In Section \ref{s3}, we derive a lower bound on the achievable ergodic secrecy rate and introduce the proposed G-NS AN precoder design. In
Section \ref{s4}, the impact of the various system and channel parameters on the secrecy performance is investigated based on the derived lower bound. In Section \ref{s5}, the
achievable secrecy rate is studied via simulation and numerical evaluation of the derived analytical expressions.  Conclusions are drawn in Section \ref{s6}.

\textit{Notation:} Superscripts $T$ and $H$ stand for the transpose and conjugate transpose, respectively. ${\bf I}_N$ is the $N$-dimensional identity matrix. The expectation operation
and the variance of a random variable are denoted by $\mathbb{E}[\cdot]$ and ${\rm var}[\cdot]$, respectively. ${\rm diag}\{{\bf x}\}$ denotes a diagonal matrix with the elements of vector
${\bf x}$ on the main diagonal. ${\rm tr}\{\cdot\}$ denotes the trace of a matrix. $\mathbb{C}^{m\times n}$ represents the space of all $m \times n$ matrices with complex-valued elements.
${\bf x} \sim \mathbb{CN}({\bf 0}_N, \boldsymbol{\Sigma})$ denotes a circularly symmetric complex Gaussian vector ${\bf x}\in \mathbb{C}^{N \times 1}$ with zero mean and covariance
matrix $\boldsymbol{\Sigma}$. $[{\bf A}]_{kl}$ denotes the element in the $k^{\rm th}$ row and $l^{\rm th}$ column of matrix ${\bf A}$. $[x]^+=\max\{x,0\}$ and $\floor{x}$ stands for the
largest integer no greater than $x$. Finally, $|{\cal S}|$ represents the cardinality of set ${\cal S}$.
\section{System and Channel Models}\label{s2}
The considered massive MIMO system model comprises an $N$-antenna BS, $K$ single-antenna mobile terminals (MTs), and an $N_E$-antenna eavesdropper. The eavesdropper is passive
in order to hide its existence from the BS and the MTs. Similar to \cite{nonideal,MRC}, we assume that after proper compensation the residual HWIs manifest themselves at the
BS and the MTs in the form of 1) multiplicative phase noises at transmitter and receiver, 2) transmit and receive power dependent distortion noises at transmitter and receiver, respectively, and 3)
amplified thermal noise at the receiver. The impact of this general HWI model on uplink training and downlink data transmission is investigated in Sections \ref{s2a} and
\ref{s2b}, respectively, and the signal model for the eavesdropper is presented in Section \ref{s2c}. {In this paper, we consider a single-cell system. This allows us to concentrate on the main focus of our work, i.e., studying the impact of HWIs on physical layer security in massive MIMO systems. Naturally, the obtained results can serve as a benchmark for multi-cell massive MIMO systems with HWIs.}

\subsection{Uplink Pilot Training under HWIs}\label{s2a}
\begin{figure}
  \centering
    \includegraphics[width=3.7in]{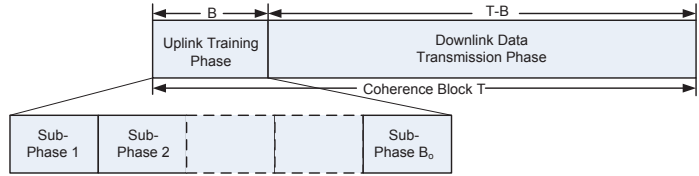}\\[-6mm]
    \caption{Uplink training and downlink transmission phase.}\label{fig01}
  \end{figure}
In massive MIMO systems, the CSI is usually acquired via uplink training by exploiting the channel reciprocity between uplink and downlink in time-division duplex (TDD) mode \cite{noncooperative,pilotcontam}. Here,
we assume that the first $B$ symbol intervals of the coherence time, which comprises $T$ symbol intervals, are used for uplink training. Thereby, we split the training phase
into $B_o$ sub-phases of lengths $B_b$, $1\le b\le B_o$, where $\sum_{b=1}^{B_o}B_b=B$, cf.~Fig.~\ref{fig01}. Furthermore, the $K$ MTs are assigned to $B_o$ disjunct sets
${\cal S}_b$, $1\le b\le B_o$, with $|{\cal S}_b| \leq B_b$ and $\sum_{b=1}^{B_o} |{\cal S}_b|=K$. In training sub-phase $b$, the MTs in set ${\cal S}_b$ emit mutually orthogonal
pilot sequences $\boldsymbol{\omega}_k=[\omega_k(1),\omega_k(2),\ldots,\omega_k(B_b)]^T \in \mathbb{C}^{B_b \times 1},k \in {\cal S}_b$, for which we assume a per-pilot
power constraint $|\omega_k(t)|^2 = p_\tau,\forall k,t$, whereas all MTs $k \notin {\cal S}_b$ are silent \footnote{{We adopt a per-pilot power constraint, as in practice, systems are peak power limited, e.g., \cite{energyspectraleff,pilotcontam,nonideal}. We note that some of our results and conclusions may change if an average power constraint for the pilot sequence was employed.}}. For larger values of $B_b$, the total energy of the pilot sequences is larger
but, as will be shown later, the loss of orthogonality caused by phase noise becomes also more pronounced. Hence, $B_b$ or equivalently $B_o$ (assuming a fixed $B$) should
be optimized for maximization of the secrecy rate. We note that the proposed pilot design is a generalization of the SO and TO pilot designs proposed in \cite{pn,MRC} which result
as special cases for $B_o=1$ and $B_o=B$, respectively.

In symbol interval $t \in {\cal T}_b$, where ${\cal T}_b$ denotes the set of symbol intervals in training sub-phase $b$, $1 \leq b \leq B_o$, the received uplink vector
${\bf y}^{\rm UL}(t) \in \mathbb{C}^{N \times 1}$ at the BS is given by
\begin{equation}\label{yULt}
{\bf y}^{\rm UL}(t)=\sum_{k \in {\cal S}_b} \boldsymbol{\Theta}_k(t){\bf g}_k (\omega_k(t)+\eta^{\rm MT}_{t,k}(t))+\boldsymbol{\eta}^{\rm BS}_r(t)+\boldsymbol{\xi}^{\rm UL}(t) .
\end{equation}
Here, the channel vector of the $k^{\rm th}$ MT, ${\bf g}_k \sim \mathbb{CN}({\bf 0}_N, \beta_k{\bf I}_N)$, is modelled as block Rayleigh fading, where $\beta_k$ denotes the
path-loss. Thereby, ${\bf g}_k$ is assumed to be constant during coherence time $T$ and change independently afterwards. In (\ref{yULt}), the terms
$\boldsymbol{\Theta}_k(t)$, $\eta^{\rm MT}_{t,k}(t)$, $\boldsymbol{\eta}^{\rm BS}_r(t)$, and $\boldsymbol{\xi}^{\rm UL}(t)$ characterize the HWIs affecting the uplink
training phase and are explained in detail in the following:\\
\textbf{1) Phase noise:} Matrix
 \begin{equation}
\boldsymbol{\Theta}_k(t)={\rm diag}\left(e^{j \theta^1_k(t)} {\bf 1}_{1 \times N/N_o},\ldots,e^{j \theta^{N_o}_k(t)}{\bf 1}_{1 \times N/N_o}\right) \in \mathbb{C}^{N \times N}
\end{equation}
models the phase noise originating from the free-running LOs equipped at the BS and the MTs \cite{pn}. Thereby, we assume that at the BS each group of $N/N_o \in \mathbb{Z}$
antennas is connected to one free-running LO. $\theta^l_k(t)=\psi_l(t)+\phi_k(t)$ is the phase noise that distorts the link between the $l^{\rm th}$ LO at the BS and the
$k^{\rm th}$ MT. Adopting the discrete-time Wiener phase noise model \cite{pn}, in time interval $t$, the phase noises at the $l^{\rm th}$ LO of the BS and the $k^{\rm th}$ MT
are modelled as $\psi_l(t) \sim \mathbb{CN}(\psi_l(t-1), \sigma^2_{\psi})$, $1 \leq l \leq N_o$, and $\phi_k(t) \sim \mathbb{CN}(\phi_k(t-1), \sigma^2_{\phi})$, $1 \leq k \leq K$, where
$\sigma^2_{\psi}$ and $\sigma^2_{\phi}$ are the phase noise (increment) variances at the BS and the MTs, respectively.\\
\textbf{2) Distortion noise:} $\eta_{t,k}^{\rm MT}(t) \in \mathbb{C}$ and $\boldsymbol{\eta}_r^{\rm BS}(t) \in \mathbb{C}^{N \times 1}$ model the additive distortion noises at the
$k^{\rm th}$ MT and the BS, respectively, which originate from residual effects after compensation of HWIs such as power amplifier non-linearities at the transmitter,
quantization noise in the analog-to-digital converters (ADCs) at the receiver, etc. \cite{nonideal}. Distortion noise is modeled as a Gaussian distributed random process in the literature
\cite{aaa,nonideal}. This model has been experimentally verified in \cite{experiment}. Furthermore, at each antenna, the distortion noise power is proportional to the corresponding
signal power, i.e., $\eta^{\rm MT}_{t,k}(t) \sim \mathbb{CN}(0,\upsilon_{t,k}^{\rm MT})$ and $\boldsymbol{\eta}^{\rm BS}_r(t) \sim \mathbb{CN}({\bf 0}_N,\boldsymbol{\Upsilon}_r^{\rm BS})$, where
\begin{equation}
\upsilon_{t,k}^{\rm MT}=\kappa^{\rm MT}_t \mathbb{E}[|\omega_k(t)|^2] \quad {\rm and} \quad \boldsymbol{\Upsilon}_r^{\rm BS}=\kappa^{\rm BS}_r \sum_{k=1}^K \mathbb{E}[|\omega_k(t)|^2]
{\bf R}^{\rm diag}_k.
\end{equation}
Here, ${\bf R}^{\rm diag}_k={\rm diag}\left(|g_k^1|^2,\ldots,|g_k^N|^2\right)$, where $g_k^i$ denotes the $i^{\rm th}$ element of ${\bf g}_k$, and parameters $\kappa^{\rm MT}_t,\kappa^{\rm BS}_r>0$
denote the ratio between the additive distortion noise variance and the signal power and are measures for the severity of the residual HWIs.\\
\textbf{3) Amplified thermal noise:} $\boldsymbol{\xi}^{\rm UL}(t)\sim \mathbb{CN}({\bf 0}_N,\xi^{\rm UL}{\bf I}_N)$ models the thermal noise amplified by the low noise amplifier and other
components such as mixers at the receiver \cite{MRC}. Therefore, the variance of this noise is generally larger than that of the actual thermal noise $\sigma_n^2$, i.e., $\xi^{\rm UL} >
\sigma^2_n$.

For channel estimation, we collect the signal vectors received during the $b^{\rm th}$ training phase in vector $\boldsymbol{\psi}_b=[({\bf y}^{\rm UL}(\overline{B}_{b-1}+1))^T,$ $\ldots,({\bf y}^{\rm UL}(\overline{B}_b))^T]^T
\in \mathbb{C}^{B_b N \times 1}$, $b=1,\ldots,B_o$, where $\overline{B}_b \triangleq  \sum_{i=1}^b B_i$ and $\overline{B}_0=0$, and define the effective channel vector at time $t$ as ${\bf g}_k(t)=\boldsymbol{\Theta}_k(t){\bf g}_k$. With these definitions,
the linear minimum mean-square error (LMMSE) estimate of the channel of MT $k \in {\cal S}_b$ at time $t \in \{B+1,\ldots, T\}$ (i.e., during the data transmission phase) can be written as \cite{MRC}
\begin{equation}\label{hatgkt1}
\hat{\bf g}_k(t)=\mathbb{E}[{\bf g}_k(t) \boldsymbol{\psi}_b^H]\left(\mathbb{E}[\boldsymbol{\psi}_b\boldsymbol{\psi}_b^H]\right)^{-1}\boldsymbol{\psi}_b=\left(\beta_k \boldsymbol{\omega}^H_k \boldsymbol{\Theta}^b_{\sigma(t)} \boldsymbol{\Sigma}_b^{-1} \otimes {\bf I}_N\right) \boldsymbol{\psi}_b,
\end{equation}
where
\begin{equation}\label{thetasigma}
\boldsymbol{\Theta}^b_{\sigma(t)}={\rm diag}\left(e^{-\frac{\sigma^2_\psi+\sigma^2_\phi}{2}|t-\overline{B}_{b-1}-1|},\ldots,e^{-\frac{\sigma^2_\psi+\sigma^2_\phi}{2}|t-\overline{B}_{b}|}\right) ~{\rm and}~ \boldsymbol{\Sigma}_b=\sum_{k \in {\cal S}_b} \beta_k \left({\bf W}_k^b+{\bf U}^b_k\right)+\xi^{\rm UL} {\bf I}_{B_b}.
\end{equation}
Here, we adopted the definitions $[{\bf W}_k^b]_{i,j}=\omega_k(i)\omega^*_k(j) e^{-\frac{\sigma^2_\psi+\sigma^2_\phi}{2}|i-j|}$, $i,j \in \{1,\ldots B_b\}$, and ${\bf U}^b_k=(\kappa^{\rm MT}_t+
\kappa^{\rm BS}_r)p_\tau {\bf I}_{B_b}$.

Considering the properties of LMMSE estimation, the channel can be decomposed as ${\bf g}_k(t)=\hat{\bf g}_k(t)+{{{\bf e}_k(t)}}$, $t=1,\ldots,B$, where $\hat{\bf g}_k(t)$ denotes the
LMMSE channel estimate given in (\ref{hatgkt1}) and ${\bf e}_k(t)$ represents the estimation error. $\hat{\bf g}_k(t)$ and ${{{\bf e}_k(t)}}$ are mutually uncorrelated and have zero mean
\cite[{Theorem 1}]{nonideal}. The error covariance matrix is given by
\begin{equation}\label{error}
\mathbb{E}[{\bf e}_k(t){\bf e}^H_k(t)]=\beta_k \left(1-\beta_k \boldsymbol{\omega}_k^H \boldsymbol{\Theta}^b_{\sigma(t)} \boldsymbol{\Sigma}_b^{-1} \boldsymbol{\Theta}^b_{\sigma(t)} \boldsymbol{\omega}_k\right) {\bf I}_N.
\end{equation}
Eqs.~(\ref{hatgkt1})-(\ref{error}) reveal that for $ |{\cal S}_b| > 1$ and $\sigma^2_\psi, \sigma^2_\phi >0$, the channel estimate of the $k^{\rm th}$ MT contains contributions from channels of other MTs emitting their pilots in
the same training sub-phase, i.e., {the pilots interfere with each other} although the emitted pilot sequences are orthogonal. This loss of orthogonality at the receiver is introduced by the phase noise via matrices $\boldsymbol{\Theta}^b_{\sigma(t)}$
and ${\bf W}_k^b$, and can be avoided only by enforcing that in any sub-phase only one MT emits its pilots, i.e., $|{\cal S}_b| =1$, $1\leq b \leq B_o$. In particular, for the case $|{\cal S}_b| =B_b=1$,
$1\leq b \leq B_o=B$, for symbol interval $t \in \{B+1, \ldots T\}$, the LMMSE channel estimate of MT $k \in {\cal S}_b$ can be simplified to
\begin{equation}\label{hatgk}
\hat{\bf g}_{k}(t)=\frac{{{p_\tau}} \beta_k e^{-\frac{\sigma^2_\psi+\sigma^2_\phi}{2}|t-b|}}{p_\tau \beta_k(1+\kappa^{\rm MT}_t+\kappa^{\rm BS}_r)+\xi^{\rm UL}} {\bf y}^{\rm UL}(b),
\end{equation}
with ${\bf y}^{\rm UL}(t)$ given in (\ref{yULt}), i.e., $\hat{\bf g}_{k}(t)$ is not affected by the channels of other MTs despite the phase noise. The corresponding error covariance matrix simplifies to
\begin{equation}
\mathbb{E}[{{{\bf e}_k(t)}{\bf e}^H_k(t)}]=\beta_k \left(1-\frac{p_\tau\beta_k {{e^{-(\sigma^2_\psi+\sigma^2_\phi)|t-b|}}}}{p_\tau \beta_k(1+\kappa^{\rm MT}_t+\kappa^{\rm BS}_r)+\xi^{\rm UL}}\right){\bf I}_N.
\end{equation}
Eqs.~(\ref{hatgkt1}) and (\ref{hatgk}) reveal that the channel estimate depends on time $t$. As a consequence, {{ideally, the channel-dependent data and AN precoders employed for downlink transmission
should be recomputed in every symbol interval of the data transmission phase, in accordance with the corresponding channel estimate, which entails a high computational complexity.}} Therefore, in the following, we assume that data and AN precoders are computed
based on the channel estimate for one symbol interval $t_0$ (e.g.,~$t_0=B+1$) and are then employed for precoding during the entire data transmission phase, i.e., for $t \in \{B+1,\ldots,T\}$. For notational conciseness,
we denote the corresponding channel estimate by $\hat{\bf g}_{k}=\hat{\bf g}_{k}(t_0)$, $k= \{1,\ldots, K\}$.

\subsection{Downlink Data Transmission and Linear Precoding}\label{s2b}
Assuming channel reciprocity, during the downlink data transmission phase, the received signal at the $k^{\rm th}$ MT in time interval $t \in \{B+1,\ldots,T\}$ is given by
\begin{equation}\label{ykt}
y^{\rm DL}_k(t)={\bf g}^H_k \boldsymbol{\Theta}^H_k(t) ({\bf x}+\boldsymbol{\eta}^{\rm BS}_t(t))+\eta^{\rm MT}_{r,k}(t)+\xi^{\rm DL}_k(t).
\end{equation}
In (\ref{ykt}), similar to the uplink, $\boldsymbol{\eta}^{\rm BS}_t(t) \sim \mathbb{CN}({\bf 0}_N,\boldsymbol{\Upsilon}^{\rm BS}_t)$ and $\eta^{\rm MT}_{r,k}(t) \sim \mathbb{CN}(0,\upsilon^{\rm MT}_{r,k}(t))$
denote the downlink distortion noise \cite{nonideal} at the BS and the $k^{\rm th}$ MT, respectively, where
\begin{equation}\label{DLim}
\boldsymbol{\Upsilon}^{\rm BS}_t=\kappa^{\rm BS}_t {\rm diag}\left(X_{11},\ldots,X_{NN}\right) \quad {\rm and} \quad \upsilon^{\rm MT}_{r,k}(t)=\kappa^{\rm MT}_r {\bf g}^H_k(t) {\bf X} {\bf g}_k(t)
\end{equation}
with ${\bf X}=\mathbb{E}[{\bf x}{\bf x}^H]$ and $X_{ii}=[{\bf X}]_{ii},i=1,\ldots,N$. Furthermore, $\xi^{\rm DL}_k(t) \sim \mathbb{CN}(0,\xi^{\rm DL})$ represents the amplified thermal noise at the $k^{\rm th}$ MT.
For simplicity of presentation, we assume that parameters $\kappa^{\rm BS}_t $, $\kappa^{\rm MT}_r$, and $\xi^{\rm DL}$ are identical for all MTs.

The downlink transmit signal ${\bf x} \in \mathbb{C}^{N \times 1}$ in (\ref{ykt}) is modeled as
\begin{equation}\label{x}
{\bf x}=\sqrt{p}{\bf F}{\bf s}+\sqrt{q}{\bf A}{\bf z} \in \mathbb{C}^{N \times 1},
\end{equation}
where the data symbol vector ${\bf s} \in \mathbb{C}^{K \times 1}$ and the AN vector ${\bf z} \in \mathbb{C}^{L \times 1}$, $L \leq N$, are multiplied by data precoder ${\bf F} \in \mathbb{C}^{N \times K}$  and AN
precoder ${\bf A} \in \mathbb{C}^{N \times L}$, respectively. As we assume that the eavesdropper's CSI is not available at the BS, AN is injected to degrade the eavesdropper's ability to decode the data
intended for the MTs \cite{zhu,zhu2,zhou}. Thereby, it is assumed that the components of ${\bf s}$ and ${\bf z}$ are independent and identically distributed (i.i.d.) circularly symmetric complex Gaussian (CSCG)
random variables, i.e., ${\bf s} \sim \mathbb{CN}({\bf 0}_K,{\bf I}_K)$ and ${\bf z} \sim \mathbb{CN}({\bf 0}_L,{\bf I}_L)$. In (\ref{x}), $p=\phi P_T/K$ and $q=(1-\phi) P_T/L$ denote the power assigned to each
MT and each column of the AN, where $P_T$ is the total power budget and $\phi \in (0,1]$ can be used to strike a balance between data transmission and AN emission. Combining (\ref{x}) and (\ref{ykt}) we obtain
\begin{equation}\label{yDLkt}
y^{\rm DL}_k(t)=\sqrt{p}{\bf g}^H_k(t) {\bf f}_k s_k+\sum_{l \neq k}^K \sqrt{p}{\bf g}^H_k(t) {\bf f}_l s_l +\sqrt{q} {\bf g}^H_k(t) {\bf A}{\bf z}+{\bf g}^H_k(t) \boldsymbol{\eta}^{\rm BS}_t(t)+\eta^{\rm MT}_{r,k}(t)+\xi^{\rm DL}_k(t),
\end{equation}
where $s_k$ and ${\bf f}_k$ denote the $k^{\rm th}$ element of ${\bf s}$ and the $k^{\rm th}$ column of matrix ${\bf F}$, respectively.

\subsection{Signal Model of the Eavesdropper}\label{s2c}
We assume that the eavesdropper is silent during the training phase, i.e., for $t \in \{1,\ldots,B\}$, and eavesdrops the signal intended for MT $k$ during the
data transmission phase, i.e., for $t \in \{B+1,\ldots,T\}$. Let ${\bf G}_E$ denote the channel matrix between the BS and the eavesdropper with i.i.d.~zero-mean complex Gaussian elements having variance $\beta_E$,
where $\beta_E$ is the path-loss between the BS and the eavesdropper. Since the capabilities of the eavesdropper are not known at the BS, we make worst-case assumptions regarding the hardware and signal
processing capabilities of the eavesdropper with respect to communication secrecy. In particular, we assume the received signal at the eavesdropper at time $t \in \{B+1,\ldots,T\}$ can be modelled as
\begin{equation}\label{yEt}
{\bf y}_E(t)={\bf G}_E^H \boldsymbol{\Psi}^H(t) ({\bf x}+\boldsymbol{\eta}^{\rm BS}_t(t)) \in \mathbb{C}^{N_E \times 1},
\end{equation}
where $\boldsymbol{\Psi}(t)={\rm diag}\left(e^{j \psi_1(t)} {\bf 1}^T_{1 \times N/N_o},\ldots,e^{j \psi_{N_o}(t)}{\bf 1}^T_{1 \times N/N_o}\right)$. Thereby, we assumed that the eavesdropper employs high-quality
hardware such that the only HWIs are the phase noise and the additive distortion noise at the BS. Eq.~(\ref{yEt}) also implies that the thermal noise at the eavesdropper is
negligibly small \cite{zhu,zhu2,zhou}. Furthermore, we assume that the eavesdropper has perfect CSI, i.e., it can perfectly estimate the effective eavesdropper channel matrix ${\bf G}_E^H\boldsymbol{\Psi}^H(t) $, and
can perfectly decode and cancel the interference caused by all MTs except for the MT of interest \cite{zhu,zhu2,zhou}. These worst-case assumptions lead to an upper bound on the ergodic capacity of the
eavesdropper {{in time interval $t$}} given by
\begin{equation}\label{CE}
C_E{{(t)}}=\mathbb{E}[\log_2(1+\gamma_E{{(t)}})]
\end{equation}
where
\begin{equation}\label{gammaE}
{{\gamma_E(t)=p {\bf g}^k_E(t) \left({\bf G}_E^H \boldsymbol{\Psi}^H(t)(q{\bf A}{\bf A}^H+\boldsymbol{\Upsilon}^{\rm BS}_t) \boldsymbol{\Psi}(t){\bf G}_E\right)^{-1}({\bf g}^k_E(t))^H}}
\end{equation}
and ${{{\bf g}^k_E(t)={\bf f}_k^H \boldsymbol{\Psi}(t){\bf G}_E}}$. We note that since we assumed that the thermal noise at the receiver of the eavesdropper is negligible, $\gamma_E{{(t)}}$, and consequently $C_E{{(t)}}$, are independent of the path-loss
of the eavesdropper, $\beta_E$. {{Furthermore, we observe from (\ref{gammaE}) that the additive distortion noise at the BS, $\boldsymbol{\eta}^{\rm BS}_t(t)$, with co-variance matrix $\boldsymbol{\Upsilon}^{\rm BS}_t$, affects the ergodic capacity of the eavesdropper in a similar manner as the injected AN.}}
\section{Achievable Ergodic Secrecy Rate in the Presence of HWIs}\label{s3}
In this section, we analyze the achievable ergodic secrecy rate of  a massive MIMO system employing non-ideal hardware. To this end, we derive a lower bound on the achievable ergodic
secrecy rate in Section \ref{s3a}, and present an asymptotic analysis for the downlink data rate of the legitimate MTs when MF data precoding is adopted by the BS in Section \ref{s3b}. In Section \ref{s3c}, a generalized NS
AN precoder is proposed to avoid the AN leakage caused by phase noise for conventional NS AN precoding. Finally, in Section \ref{s3d}, a simple closed-form upper bound for the eavesdropper's capacity
for the new AN precoder is presented.

\subsection{Lower Bound on Achievable Ergodic Secrecy Rate}\label{s3a}
In this paper, we assume that communication delay is tolerable and coding over many independent channel realizations is possible. Hence, we adopt the ergodic secrecy rate achieved by a given MT as
performance metric \cite{zhou}.

Before analyzing the secrecy rate, we first employ \cite[{Lemma 1}]{MRC} to obtain a lower bound on the achievable rate for the multiple-input single-output (MISO) phase noise channel given by (\ref{ykt}).
In particular, the achievable rate of the $k^{\rm th}$ MT, $1 \leq k \leq K$, in symbol interval $t \in \{B+1,\ldots,T\}$ is lower bounded  by
\begin{equation}\label{Rkt}
R_k(t) \geq \underline{R}_k(t)=\log_2 (1+\gamma_k(t)),
\end{equation}
with SINR $\gamma_k(t)=$
\begin{equation}\label{gammak}
\frac{p\left|\mathbb{E}\left[{\bf g}^H_k(t) {\bf f}_k\right]\right|^2}{\sum\limits_{l=1}^K p\mathbb{E}\left[\left|{\bf g}^H_k(t) {\bf f}_l\right|^2\right]-p\left|\mathbb{E}\left[{\bf g}^H_k(t) {\bf f}_k\right]\right|^2
+\mathbb{E}\left[{\bf g}^H_k(t) (q{\bf A}{\bf A}^H+\boldsymbol{\Upsilon}^{\rm BS}_t) {\bf g}_k(t)\right]+\mathbb{E}\left[\upsilon^{\rm MT}_{k,r}(t)\right]+\xi^{\rm DL}}.
\end{equation}
The expectation operator in (\ref{gammak}) is taken with respect to channel vectors, ${\bf g}_k$, as well as the phase noise processes, $\psi_l(t)$ and $\phi_k(t)$. {The rate given in (\ref{Rkt}) is achievable because: 1) The SINR in (\ref{gammak}) is underestimated by assuming that only the BS has channel estimates, while the MTs only know the mean of the effective channel gain $\left|\mathbb{E}\left[{\bf g}^H_k(t) {\bf f}_k\right]\right|$ and employ it for signal detection. The deviation from the average effective channel gain is treated as Gaussian noise having variance $\mathbb{E}\left[\left|{\bf g}^H_k(t) {\bf f}_k\right|^2\right]-|\mathbb{E}\left[{\bf g}^H_k(t) {\bf f}_k\right]|^2$, cf.~\cite{pilotcontam,zhu,zhu2,wyp}; 2) Following \cite[Lemma 1]{MRC}, we treat the multiuser interference and distortion noises as independent Gaussian noises, which is a worst-case assumption for the calculation of the mutual information. The tightness of the bound will be confirmed in Section \ref{s5}.}
Based on (\ref{Rkt}), we provide a lower bound on the achievable ergodic secrecy rate of the $k^{\rm th}$ MT, $1 \leq k \leq K$, in the following Lemma.

\textit{Lemma 1}: The achievable ergodic secrecy rate of the $k^{\rm th}$ MT, $1\leq k \leq K$, is bounded below by
\begin{equation}\label{rseck}
{R}^{\rm sec}_k \geq \underline{R}^{\rm sec}_k =\frac{1}{T} \sum_{t \in \{B+1,\ldots,T\}} \left[\underline{R}_k(t)-C_E{{(t)}}\right]^+,
\end{equation}
where $\underline{R}_k(t)$, $1 \leq k \leq K$, is the lower bound of the achievable ergodic rate of the $k^{\rm th}$ MT given in (\ref{Rkt}) and $C_E{{(t)}}$ is the ergodic capacity between the BS and the eavesdropper given in (\ref{CE}).
\begin{IEEEproof}
Please refer to Appendix B.
\end{IEEEproof}
The sum in (\ref{rseck})
is over the $T-B$ time slots used for data transmission. Motivated by the coding scheme for the non-secrecy case in \cite{TRMRC}, a similar coding scheme that supports the secrecy rate given in (\ref{rseck})
is described as follows. For a given $t \in \{B+1,\ldots,T\}$, the statistics of ${\bf g}_k(t)$ in (\ref{gammak}) given the estimate $\hat{\bf g}_k$ are identical across all coherence intervals and the corresponding
channel realizations are i.i.d. Hence, we employ $T-B$ parallel channel codes for each MT; one code for each time $t \in \{B+1,\ldots,T\}$, i.e., the $t^{\rm th}$ channel code is employed across the $t^{\rm th}$ time slots  of multiple
coherence intervals. Then, at each MT, the $t^{\rm th}$ received symbols across the multiple coherence intervals are jointly decoded \cite{TRMRC}. With this coding strategy the ergodic secrecy rate given
in (\ref{rseck}) is achieved provided the parallel codes span sufficiently many (ideally an infinite number) of independent channel realizations ${\bf g}_k$  and phase noise samples $\psi_l(t)$ and $\phi_k(t)$.

\subsection{Asymptotic Analysis of Achievable Rate for MF Precoding}\label{s3b}
In this subsection, we analyze the lower bound on the achievable ergodic rate of the $k^{\rm th}$ MT, $1 \leq k \leq K$, in (\ref{Rkt}) in the asymptotic limit $N,K\to\infty$ for fixed ratio $\beta=K/N$. Thereby, we adopt
MF precoding at the BS, i.e., ${\bf f}_k=\hat{\bf g}_{k}/\|\hat{\bf g}_{k}\|$, as is commonly done for massive MIMO systems because of complexity concerns for more sophisticated precoder designs. In the following Lemma,
we provide a closed-form expression for the gain of the desired signal.

\textit{Lemma 2}: For MF precoding at the BS, the numerator of (\ref{gammak}) reflecting the gain of the desired signal at MT $k \in {\cal S}_b$, $1 \leq b \leq B_o$, can be expressed as
\begin{equation}\label{lambdak}
 \mathbb{E}\left[{\bf g}^H_k \boldsymbol{\Theta}^H_k(t) {\bf f}_k\right]=\sqrt{\beta_k N \lambda_k} \cdot e^{-\frac{\sigma^2_\psi+\sigma^2_\phi}{2}|t-t_0|},~ {\rm where}~
 \lambda_k=\beta_k \boldsymbol{\omega}_k^H \boldsymbol{\Theta}^b_{\sigma(t_0)} \boldsymbol{\Sigma}_b^{-1} \boldsymbol{\Theta}^b_{\sigma(t_0)} \boldsymbol{\omega}_k.
\end{equation}
\begin{IEEEproof}
Please refer to Appendix C.
\end{IEEEproof}
The term $e^{-\frac{\sigma^2_\psi+\sigma^2_\phi}{2}|t-t_0|}$ in (\ref{lambdak}) reveals the impact of the accumulated phase noise from the time of channel estimation, $t_0$, to the time of
data transmission, $t$, on the received signal strength at MT $k$. On the other hand, the phase noise within the training phase affects $\lambda_k$, and consequently the received signal strength, via
$\boldsymbol{\Theta}^b_{\sigma(t_0)}$ and $\boldsymbol{\Sigma}_b$, cf.~(\ref{thetasigma}), when multiple pilot sequences are simultaneously emitted in a given training sub-phase. In contrast,
when TO pilots are adopted, i.e., only a single user emits pilots in each training sub-phase and $B_b=1$, $1\le b\le B$, $\lambda_k$ in (\ref{lambdak}) reduces to $\lambda_k=\frac{p_\tau \beta_k}{p_\tau \beta_k(1+
\kappa^{\rm MT}_t+\kappa^{\rm BS}_r)+\xi^{\rm UL}}$ and is not affected by the phase noise.

Next, an expression for the multiuser interference power in the first term of the denominator of (\ref{gammak}) is derived.

\textit{Lemma 3}: When MF precoding is adopted at the BS, the power of the multiuser interference caused by the signal intended for the $l^{\rm th}$ MT, $l \neq k$,  at MT $k \in {\cal S}_b$, $1 \leq b \leq B_o$,
is given by
\begin{equation}\label{ak1}
\mathbb{E}\left[\left|{\bf g}^H_k \boldsymbol{\Theta}^H_k(t) {\bf f}_l\right|^2\right]=\left(\beta_k+ \left(X^{(1)}_{k,l}+X^{(2)}_{k,l}+X^{(3)}_{k,l}\right)\left(\frac{1-\epsilon}{N_o}+\epsilon\right)\right),~{\rm if}~l \in {\cal S}_b
\end{equation}
and by $\beta_k$ otherwise. Here, $\epsilon=e^{-\sigma^2_\psi | t-t_0|}$, $X^{(1)}_{k,l}=\frac{\beta^2_k \boldsymbol{\omega}^H_l \boldsymbol{\Theta}^b_{\sigma(t_0)}
\boldsymbol{\Sigma}_b^{-1}{\bf U}^b_k\boldsymbol{\Sigma}_b^{-1}\boldsymbol{\Theta}^b_{\sigma(t_0)}\boldsymbol{\omega}_l}{\boldsymbol{\omega}_l^H
\boldsymbol{\Theta}^b_{\sigma(t_0)} \boldsymbol{\Sigma}_b^{-1} \boldsymbol{\Theta}^b_{\sigma(t_0)} \boldsymbol{\omega}_l}$,
$X^{(2)}_{k,l}=\frac{N}{N_o} \cdot \frac{\beta^2_k \boldsymbol{\omega}^H_l \boldsymbol{\Theta}^b_{\sigma(t_0)}\boldsymbol{\Sigma}_b^{-1}{\bf W}_k^b\boldsymbol{\Sigma}_b^{-1}
\boldsymbol{\Theta}^b_{\sigma(t_0)}\boldsymbol{\omega}_l}{\boldsymbol{\omega}_l^H \boldsymbol{\Theta}^b_{\sigma(t_0)} \boldsymbol{\Sigma}_b^{-1}
\boldsymbol{\Theta}^b_{\sigma(t_0)} \boldsymbol{\omega}_l}$, and $X^{(3)}_{k,l}=N \left(1-\frac{1}{N_o}\right) \cdot \frac{\bigg|\beta_k \boldsymbol{\omega}_k^H
\boldsymbol{\Theta}^b_{\sigma(t_0)} \boldsymbol{\Sigma}_b^{-1} \boldsymbol{\Theta}^b_{\sigma(t_0)} \boldsymbol{\omega}_l\bigg|^2}{\boldsymbol{\omega}_l^H
\boldsymbol{\Theta}^b_{\sigma(t_0)} \boldsymbol{\Sigma}_b^{-1} \boldsymbol{\Theta}^b_{\sigma(t_0)} \boldsymbol{\omega}_l}$.
\begin{IEEEproof}
Please refer to Appendix D.
\end{IEEEproof}
Lemma 3 confirms that when the number of BS antennas is sufficiently large, i.e., $N \to \infty$, as long as $l \notin {\cal S}_b$, the impact of the multiuser interference from the $l^{\rm th}$ MT vanishes,
as is commonly assumed in the massive MIMO literature, e.g.~\cite{noncooperative,energyspectraleff}. However, the same is not true for MTs that emit pilots in the same training sub-phase
as MT $k$, i.e., MTs $l \in {\cal S}_b$. Because of the impairment incurred by the phase noise during the training phase, the interference power of these MTs grows linearly with $N$ and does
not vanish compared to the strength of the desired signal in (\ref{lambdak}) in the limit of $N\to\infty$.

Furthermore, for the summand with $l=k$ in the sum in the first term of the denominator of (\ref{gammak}), we obtain $\mathbb{E}\left[\left|{\bf g}^H_k \boldsymbol{\Theta}^H_k(t) {\bf f}_k\right|^2\right]=$
\begin{equation}\label{exs}
\mathbb{E}\left[{\rm tr}\left({\bf g}_{k}(t_0) {\bf g}^H_{k}(t_0) \boldsymbol{\Psi}^H_{t_0}(t)
\frac{\hat{\bf g}_{k}\hat{\bf g}^H_k}{\|\hat{\bf g}_{k}\|^2}\boldsymbol{\Psi}_{t_0}(t)\right)\right]=\beta_k +\beta_k (N-1) \lambda_k \left(\frac{1-\epsilon}{N_o}+\epsilon\right),
\end{equation}
where $k \in {\cal S}_b$ and $\boldsymbol{\Psi}_{t_0}(t)$ is defined in Appendix C. The last equality in (\ref{exs}) is obtained by applying Theorem 1 in Appendix A \cite{free}. The variance of the gain of the desired signal, ${\bf g}^H_k \boldsymbol{\Theta}^H_k(t) {\bf f}_k$, is obtained by subtracting the right hand side
of (\ref{exs}) from the square of the right hand side of (\ref{lambdak}).

The two terms in the denominator of (\ref{gammak}) originating from the HWIs at the BS and the $k^{\rm th}$ MT, i.e., $\boldsymbol{\eta}^{\rm BS}_t(t)$ and $\eta^{\rm MT}_{r,k}(t)$, respectively,
can be calculated as
\begin{equation}
\label{eq21}
\mathbb{E}\left[\left|{\bf g}^H_k \boldsymbol{\Theta}^H_k(t)\boldsymbol{\Upsilon}^{\rm BS}_t \boldsymbol{\Theta}_k(t) {\bf g}_k\right|\right]=\beta_k \kappa^{\rm BS}_t P_T \quad {\rm and}
\quad \mathbb{E}\left[\upsilon^{\rm MT}_{r,k}(t)\right]=\beta_k \kappa^{\rm MT}_r P_T.
\end{equation}
Substituting the results in (\ref{lambdak})-(\ref{eq21}) into (\ref{gammak}), we obtain the received SINR at MT $k \in {\cal S}_b$ in symbol interval $t$ as
\begin{equation}\label{gammakt}
\gamma_k(t)=\frac{p N \beta_k \overline{\lambda}_k}{p \beta_k ({a}_{k}+c_k)+q  L^k_{\rm AN}+\beta_k (\kappa^{\rm BS}_t+\kappa^{\rm MT}_r)P_T+\xi^{\rm DL}},
\end{equation}
with
\begin{equation}\label{ak}
a_k=\sum_{l\in {\cal S}_b}\left(1+ \left(X^{(1)}_{k,l}+X^{(2)}_{k,l}+X^{(3)}_{k,l}\right)\left(\frac{1-\epsilon}{N_o}+\epsilon\right)/\beta_k\right)+(K-|{\cal S}_b|),
\end{equation}
\begin{equation}\label{ck}
c_k=\left(1-\frac{1}{N_o}\right)(1-\epsilon)+[(N-1) \lambda_k+1] \left(\frac{1-\epsilon}{N_o}
+ \epsilon\right)-N \overline{\lambda}_k,
\end{equation}
where $\overline{\lambda}_k=\lambda_k e^{-(\sigma^2_\psi+\sigma^2_\phi)|t-t_0|}$. Furthermore, $a_k$ and $c_k$ represent the multiuser interference received at the $k^{\rm th}$ MT
and the variance of the gain of the desired signal, respectively. Moreover, the term $L^k_{\rm AN}=\mathbb{E}\left[{\bf g}^H_k \boldsymbol{\Theta}^H_k(t) {\bf A}{\bf A}^H
\boldsymbol{\Theta}_k(t) {\bf g}_k\right]$ in (\ref{gammakt}) represents the AN leakage in the received signal of the $k^{\rm th}$ MT in time slot $t$. This term will be
characterized in detail for the considered AN precoders in Section \ref{s3c}.

\subsection{Generalized NS AN Precoding}\label{s3c}
The AN leakage term $L^k_{\rm AN}$ in (\ref{gammakt}) depends on the particular AN precoder used. Therefore, in this subsection, we first evaluate $L^k_{\rm AN}$ for the conventional NS precoder, where
${\bf A}$ is designed to lie in the NS of the estimated channel vectors of all MTs, $\hat{\bf g}_{k}$, $1 \leq k \leq K$, which is the most common design used in the literature \cite{zhu,zhu2,zhou}. Subsequently,
we propose and analyze the G-NS AN precoder design which is less sensitive to HWIs than the conventional NS design.

The AN leakage incurred by the conventional NS AN precoder is given in the following Lemma.

\textit{Lemma 4}: For the conventional NS AN precoder, where $L=N-K$ \cite{zhu,zhu2,zhou}, the AN leakage power received at MT $k \in {\cal S}_b$ in time interval $t$ is given by
\begin{equation}\label{AN1}
L^k_{\rm AN}=\beta_k (N-K)\left(\left(1-\frac{1}{N_o}\right)\left(1-\epsilon\right)+1-\lambda_k\right).
\end{equation}
\begin{IEEEproof}
Please refer to Appendix E.
\end{IEEEproof}
In Lemma 4, the terms $\epsilon$ and $\lambda_k$ reflect the negative impact of the HWIs on the AN power leakage. If only one LO is employed, i.e., $N_o=1$, the impact of $\epsilon$ is eliminated.
However, the negative effect of $\epsilon$ increases as the number of LOs, $N_o$, increases since the phase noise processes of different LOs are independent destroying the orthogonality of the columns of
${\bf A}$ and ${\bf g}_{k}(t)$, $1 \leq k \leq K$.

This problem can be mitigated by employing $M_o$ NS AN precoders where each precoder encodes the data signals intended for the antennas connected to $N_o/M_o$ LOs. Thereby, $N_o$ is assumed to be
a multiple of $M_o$, i.e., $N_o/M_o \in \mathbb{Z}$. The resulting AN preorder is referred to as G-NS AN precoder. More in detail, for the G-NS AN precoder, we divide each channel estimation vector,
$\hat{\bf g}_{k}$, $1 \leq k \leq K$, into $M_o$ sub-vectors
\begin{equation}
\hat{\bf g}_{k}=\bigg[\left(\hat{\bf g}^{(1)}_k\right)^T,\left(\hat{\bf g}^{(2)}_k\right)^T,\ldots,\left(\hat{\bf g}^{(M_o)}_k\right)^T\bigg]^T,
\end{equation}
where $\hat{\bf g}^{(m)}_k \in \mathbb{C}^{N/M_o \times 1}$, which contains the $((m-1)N/M_o+1)^{\rm th}$ to the $(m N/M_o)^{\rm th}$ elements of $\hat{\bf g}_{k}$ for
$1 \leq m \leq M_o$. Correspondingly, we split matrix ${\bf A}$ into $M_o$ sub-matrices as follows
\begin{equation}
{\bf A}=\bigg[{\bf A}^T_{(1)},{\bf A}^T_{(2)}\ldots,{\bf A}^T_{(M_o)}\bigg]^T,
\end{equation}
with ${\bf A}_{(m)} \in \mathbb{C}^{N/M_o \times (N/M_o-K)}$, $1 \leq m \leq M_o$, i.e., we have $L=N/M_o-K$. Now, matrix ${\bf A}_{(m)}$ is designed to lie in the null-space of $\hat{\bf g}^{(m)}_k$, $1 \leq k \leq K$,
i.e., ${\bf A}_{(m)}\hat{\bf g}^{(m)}_k={\bf 0}$, $1 \leq k \leq K$, $1 \leq m \leq M_o$. For $M_o=1$, the G-NS precoder simplifies to the conventional NS precoder. On the other hand, for $M_o=N_o$, the antennas connected
to each LO have their own NS AN precoder.

The AN leakage of the G-NS precoder is analyzed in the following Lemma.

\textit{Lemma 5}: For the G-NS AN precoder, where $L=N/M_o-K$ and $1\le M_o\le N_o$, the AN leakage power received at MT $k \in {\cal S}_b$ in time interval $t$ is given by
\begin{equation}\label{AN2}
L^k_{\rm AN}=\beta_k \left(\frac{N}{M_o}-K\right) \left(\left(1-\frac{M_o}{N_o}\right)\left(1-\epsilon\right)+1-\lambda_k\right).
\end{equation}
\begin{IEEEproof}
Please refer to  Appendix F.
\end{IEEEproof}
Several observations can be made from (\ref{AN2}). First, we note that, as expected, for $M_o=1$, (\ref{AN2}) reduces to (\ref{AN1}). Second, the negative impact of the phase noise via $\epsilon$ on the AN leakage
can be completely eliminated by choosing $M_o=N_o$. Third, the G-NS precoder requires the calculation of $M_o$ null spaces of dimension $N/M_o\times K$. Hence, the computational complexity increases with $M_o$.
We will elaborate on the optimal choice of $M_o$ in Sections \ref{s4} and \ref{s5}.

{\textit{Remark 1}: We note that the proposed G-NS AN precoder is not optimal for the maximization of the achievable secrecy rate. Nevertheless, the G-NS AN precoder achieves high performance and facilitates the derivation of closed-form expressions for the achievable secrecy rate, which provide substantial technical insight for system design. Optimizing the AN precoder for maximization of the secrecy rate \cite{optAN} is an interesting topic for future work.}

The achievable rates of MT $k \in {\cal S}_b$ in time slot $t$ with conventional NS and G-NS precoding are obtained by inserting (\ref{AN1}) and (\ref{AN2}) into (\ref{gammakt}), respectively. Hence, for the proposed
G-NS precoder, we obtain
\begin{equation}\label{Rkunder}
\underline{R}_k(t) =\log_2 \left(1+\frac{\overline{\lambda}_k \phi N}{ (a_k+c_k-\beta \mu_k) \phi + \beta \mu_k +\xi_k}\right),
\end{equation}
where $\mu_k=(\frac{N}{M_o}-K)\left(\left(1-\frac{M_o}{N_o}\right)\left(1-\epsilon\right)+1-\lambda_k\right)$, $\xi_k=\beta(\kappa^{\rm MT}_r+\kappa^{\rm BS}_t+\xi^{\rm DL}/(\beta_k P_T))$, and $\beta=K/N>0$.

\subsection{Upper Bound on the Eavesdropper's Capacity}\label{s3d}
In the following Proposition, we provide a tight and tractable upper bound on eavesdropper's capacity.

\textit{Proposition 1}: For $N \to \infty$ and (G-)NS AN precoding, the eavesdropper's capacity in (\ref{CE}) can be  upper bounded as
\begin{equation}\label{Cup1}
C_E{{(t)}} \leq \overline{C}_E=\log_2 \left(1+ \frac{p N_E}{q L +\kappa^{\rm BS}_t P_T- \chi N_E}\right),~ {\rm with} ~
\chi=\frac{(1+\kappa^{\rm BS}_t)^2 q^2 L+(\kappa^{\rm BS}_t)^2 p^2 K }{(1+\kappa^{\rm BS}_t) q L+\kappa^{\rm BS}_t p K},
\end{equation}
for $q L +\kappa^{\rm BS}_t P_T > \chi N_E$, and where $L=N-K$ and $L=N/M_o-K$ for the conventional NS and the G-NS precoders, respectively.
\begin{IEEEproof}
{Please refer to Appendix G.}
\end{IEEEproof}
We observe from (\ref{Cup1}) that, as expected, the capacity of the eavesdropper is increasing in the number of its equipped antennas, $N_E$. {{Another non-trivial observation is that the bound provided in Proposition 1 is no longer a function of time slot index $t$, due to the worst-case assumption that the eavesdropper has perfect instantaneous CSI, cf. Section \ref{s2c}.}} Interestingly, when no AN is injected, i.e., $q=0$,
(\ref{Cup1}) reduces to
\begin{equation}\label{Cup1q}
\overline{C}_E\bigg|_{q=0}=\log_2 \left(1+\frac{N_E}{\kappa^{\rm BS}_t (K- N_E)}\right),
\end{equation}
for $K > N_E$. For perfect BS hardware, we have $\kappa^{\rm BS}_t \to 0$ and $\overline{C}_E\to \infty$ making secure communication impossible. Hence, if AN is not injected, HWIs may in fact be
beneficial for secure communication as the distortion noise at the BS acts like AN and may facilitate secrecy. This surprising insight will be studied more carefully in the next section. Furthermore, the
number of independent distortion noise processes at the BS is equal to the number of users, $K$. Hence, $K>N_E$ is needed to prevent the eavesdropper from nulling out the distortion noise and for achieving
secrecy.

{The worst-case ergodic secrecy rate achieved by MT $k$ in time slot $t$ is lower bounded by $\bigg[\underline{R}_k(t)-\overline{C}_E\bigg]^+$, where $\underline{R}_k(t)$ and $\overline{C}_E$ are given in (\ref{Rkunder}) and (\ref{Cup1}), respectively. Hence, although the CSI and location information of the eavesdropper assumed to be not available at the BS, with the proposed transmission strategy, the BS can still guarantee the derived worst-case ergodic secrecy rate in the presence of HWIs. In non-worst-case scenarios, higher ergodic secrecy rates are expected. }
\section{Guidelines for System Design}\label{s4}
In this section, we exploit the analytical results derived in the previous section to gain some insight into the impact of the various system and HWI parameters on system design. To this end, we carefully study the
closed-form lower bound on the achievable ergodic secrecy rate obtained by combining (\ref{rseck}), (\ref{Rkunder}), and (\ref{Cup1}).

\subsection{Design of the Pilot Sequences}\label{s4a}
Assuming that we assign the maximum number of users to each training sub-phase, i.e., $|{\cal S}_b| = B_b$, the relevant design parameter for the pilot sequences is the number of training sub-phases $B_o$,
or equivalently, the size of the training sub-phases $B_b$ as $\sum_{b=1}^{B_o}B_b=B$. In particular, $B_b$ affects the lower bound on the achievable ergodic rate of MT $k$ in (\ref{Rkunder}) via $\lambda_k$,
$a_k$, and $c_k$, where $c_k$ becomes proportional to $\lambda_k$ for $N\to \infty$, cf.~(\ref{ck}). Thereby, close inspection of (\ref{lambdak}) reveals that $\lambda_k$,  which reflects the power of the received
useful signal, is not monotonic in $B_b$. This can be explained as follows. On the one hand, since the power of each pilot symbol is constrained, i.e., $|\omega_k(t)|^2 =p_\tau$, $\forall k,t$, the sum power of the
pilot sequence per MT increases with $B_b$. On the other hand, for larger $B_b$, more MTs are allowed to emit pilots in training sub-phase $b$ introducing more {{mutual pilot interference}} due to phase noise. This has an
adverse effect on the quality of the channel estimate and consequently on the power of the received useful signal. Similarly, close inspection of (\ref{ak}) reveals that $a_k$, which reflects the multiuser interference
incurred to the $k^{\rm th}$ MT, is a monotonically increasing function of $B_b$, as a lower channel estimation accuracy {{caused by more mutual pilot interference,}} gives rise to stronger multiuser interference. Considering the behaviour of $\lambda_k$, $a_k$,
and $c_k$ and their impact on the achievable ergodic rate of MT $k$ in (\ref{Rkunder}), we conclude that $B_b$, $1 \leq b \leq B_o$, should be optimized and the optimal value depends on the channel and HWI
parameters. Thereby, the optimal $B_b$ is decreasing in the phase noise variances, $\sigma_\psi^2$ and $\sigma_\phi^2$, as the degradation introduced by concurrent pilot emission by multiple MTs
is increasing in these parameters. This conclusion will be verified in Section \ref{s5b} by numerically evaluating (\ref{rseck}).

\subsection{Selection of $M_o$ for G-NS AN Precoding}\label{s4b}
The number of G-NS AN precoding sub-matrices, $M_o$, $1\le M_o \le N_o$, employed affects the achievable ergodic secrecy rate via the AN leakage $L_{\rm AN}^k$ in (\ref{AN2}) and via the (bound on the)
eavesdropper capacity $\overline{C}_E$ in (\ref{Cup1}). The AN leakage is a decreasing function with respect to $M_o$, i.e., as far as the AN leakage is concerned, $M_o=N_o$ is preferable. On the other hand,
since the dimensionality of the G-NS AN precoder is given by $L=N/M_o-K$, the eavesdropper capacity is an increasing function of $M_o$, cf.~(\ref{Cup1}), which has a negative effect on the ergodic secrecy rate.
Hence, $M_o$ has to be optimized. Since the eavesdropper capacity does not depend on the phase noise, we expect that the optimal $M_o$ increases with increasing BS phase noise variance, $\sigma_\psi^2$,
as $\sigma_\psi^2$ affects the AN leakage via $\epsilon$ in (\ref{AN2}). This conjecture will be numerically verified in Section \ref{s5d}.

\subsection{Secrecy in the Absence of AN}\label{s4c}
In \cite{zhu,zhu2} it was shown that if perfect hardware is employed, injection of AN is necessary to achieve secrecy. In particular, without AN generation, under worst-case assumptions regarding the noise at the
eavesdropper, the eavesdropper capacity is unbounded. On the other hand, we showed in Section \ref{s3d} that in the presence of HWIs the eavesdropper capacity is bounded since the distortion noise generated
at the BS has a similar effect as AN. Motivated by this observation, in this section, we calculate the maximum number of eavesdropper antennas $N_E$ that can be tolerated if a positive secrecy rate is desired without
AN emission.

If AN is not emitted, we have $\phi=1$ or $q=0$. In this case, the proposed lower bound on the ergodic secrecy rate of the $k^{\rm th}$ MT in time interval $t$ simplifies to
\begin{equation}\label{Rsecq}
\underline{R}^{\rm sec}_k(t) \bigg|_{q=0}=\bigg[\log_2 \left(1+\frac{\overline{\lambda}_k N}{a_k +c_k+\xi_k}\right)-\log_2 \left(1+\frac{\alpha}{\kappa^{\rm BS}_t (\beta- \alpha)}\right)\bigg]^+.
\end{equation}
where $\alpha=N_E/N$ denotes the normalized number of eavesdropper antennas. In the following Proposition, we provide a condition for the number of eavesdropper antennas that has to be met for secure
communication to be possible.

\textit{Proposition 2}: If AN is not generated, the maximum number of eavesdropper antennas that the $k^{\rm th}$ MT can tolerate while achieving a positive ergodic secrecy rate is
$N_E=\floor{\alpha_{\rm AN}N}$, where
\begin{equation}\label{alphaAN}
\alpha_{\rm AN}= \frac{\overline{\lambda}_k N \kappa^{\rm BS}_t \beta}{\overline{\lambda}_k N \kappa^{\rm BS}_t +a_k+c_k+\xi_k}\Big|_{t=B+1}.
\end{equation}
\begin{IEEEproof}
First, we note that $\underline{R}_k(t)$ is a decreasing function of $t$. Hence, considering (\ref{rseck}), it is sufficient to ensure $\underline{R}_k(B+1)>\overline{C}_E$ for achieving a positive
ergodic secrecy rate. Eq.~(\ref{alphaAN}) is obtained by setting (\ref{Rsecq}) to zero and observing that $\underline{R}^{\rm sec}_k(t) \bigg|_{q=0}$ is a decreasing function of $\alpha$.
\end{IEEEproof}
Eq.~(\ref{alphaAN}) clearly shows that the additive distortion noise at the BS is essential for achieving a positive secrecy rate if AN is not injected as $\alpha_{\rm AN}=0$ results if $\kappa^{\rm BS}_t =0$.
On the other hand, $\alpha_{\rm AN}$ is a decreasing function of all other HWI parameters, i.e., $\kappa^{\rm BS}_r$, $\kappa^{\rm MT}_t$, $\kappa^{\rm MT}_r$, $\xi^{\rm DL}$, $\sigma_\psi^2$, and
$\sigma_\psi^2$, as the corresponding HWIs affect only the achievable ergodic rate of the MT but not the ergodic capacity of the eavesdropper. We note that $\alpha_{\rm AN}$ is an increasing function of
$\beta$ since the dimensionality of the additive distortion noise at the BS is proportional to $\beta$.

\subsection{Maximum Number of Eavesdropper Antennas}\label{s4d}
Now, we consider the maximum number of eavesdropper antennas that can be tolerated if a positive ergodic secrecy rate is desired and AN injection is possible. Combining (\ref{rseck}), (\ref{Rkunder}), and (\ref{Cup1}),
the lower bound on the ergodic secrecy rate in time interval $t$ can be expressed as
\begin{equation}\label{Rsec}
\underline{R}^{\rm sec}_k(t) =\bigg[\log_2 \left(1+\frac{\overline{\lambda}_k \phi N}{(a_k+c_k) \phi+\beta \mu_k(1-\phi) +\xi_k}\right) -\log_2 \left(1+\frac{ \alpha \phi}{\beta (1-\phi+\kappa^{\rm BS}_t-\chi' \alpha)}\right)\bigg]^+,
\end{equation}
where $\chi'=\frac{(1+\kappa^{\rm BS}_t)^2 (1-\phi)^2N/L+(\kappa^{\rm BS}_t)^2 \phi^2/\beta}{1-\phi+\kappa^{\rm BS}_t}$.

\textit{Proposition 3}: If AN injection is possible, a positive secrecy rate can be achieved by the $k^{\rm th}$ MT if the number of eavesdropper antennas does not exceed $N_E=\floor{\alpha_{\rm sec}N}$, where
\begin{equation}\label{alphasec}
\alpha_{\rm sec}=\frac{(1+\kappa^{\rm BS}_t) \overline{\lambda}_k L }{L/N( \mu_k+\kappa^{\rm MT}_r+\kappa^{\rm BS}_t+\xi^{\rm DL}/(\beta_k P_T))+ \overline{\lambda}_k N(1+\kappa^{\rm BS}_t)}\Big|_{t=B+1}
\end{equation}
and $\phi\to 0$, i.e., almost all transmit power is employed for AN generation.
\begin{IEEEproof}
Exploiting again that $\underline{R}_k(t)$ is a decreasing function of $t$ it suffices to consider the ergodic secrecy rate for $t=B+1$. Then, an expression for $\alpha_{\rm sec}$ is obtained by setting
$\underline{R}^{\rm sec}_k(t)$ in (\ref{Rsec}) to zero. This expression is monotonically decreasing in $\phi$ and hence can be further simplified by letting $\phi\to 0$ which yields (\ref{alphasec}).
\end{IEEEproof}
Proposition 3 reveals that, as expected, the number of eavesdropper antennas that can be tolerated increases with the channel estimation accuracy (i.e., $\overline{\lambda}_k$) and the
number of spatial dimensions available for AN (i.e., $L$). Furthermore, similar to $\alpha_{\rm AN}$, $\alpha_{\rm sec}$ is a decreasing function of the HWI parameters $\kappa^{\rm BS}_r$, $\kappa^{\rm MT}_t$,
$\kappa^{\rm MT}_r$, $\xi^{\rm DL}$, $\sigma_\psi^2$, and $\sigma_\psi^2$, and an increasing function of $\kappa^{\rm BS}_t$. However, unlike $\alpha_{\rm AN}$, $\alpha_{\rm sec}$ is independent
of $\beta$.

\subsection{Number of LOs}\label{s4f}
The number of LOs, $N_o$, affects the ergodic secrecy rate via the terms $a_k$, $c_k$, and $\mu_k$ in the achievable ergodic rate in (\ref{Rkunder}). For $N\to\infty$, $a_k$ and $c_k$ are decreasing
functions of $N_o$, i.e., less multiple access interference is caused if more LOs are employed, whereas the AN leakage term $\mu_k$ is an increasing function in $N_o$. Therefore, considering the specific form of the
denominator of the fraction inside the logarithm in (\ref{Rkunder}), the optimal value of $N_o$, which maximizes the ergodic secrecy rate, depends on $\phi$. In particular, for a given $M_o$, for
$\phi=1$ no AN is injected and $\mu_k$ cancels in the expression for the achievable ergodic rate in (\ref{Rkunder}). Hence, in this case, the ergodic secrecy rate is a monotonically increasing
function of $N_o$, i.e., increasing the number of LOs is beneficial. On the other hand, for a given $M_o$, for $\phi<1$, the optimal $N_o$ maximizing the ergodic secrecy rate can be found by performing a
numerical search based on (\ref{Rkunder}).

We note that by employing G-NS AN generation and enforcing $M_o=N_o$, we can avoid the harmful effect of the multiple LOs on the AN leakage term $\mu_k$. In this case, the achievable ergodic
rate of the MT becomes an increasing function of $M_o=N_o$. However, at the same time, the number of dimensions available for AN injection, $L=N/M_o-K$, is a decreasing function of $M_o=N_o$.
Therefore, the optimal $M_o=N_o$ maximizing the ergodic secrecy rate has to be found again by a numerical search.

\subsection{Are HWIs Beneficial for Security?}\label{s4g}
Since the HWI parameters $\kappa^{\rm BS}_r$, $\kappa^{\rm MT}_t$, $\kappa^{\rm MT}_r$, $\xi^{\rm DL}$, $\sigma_\psi^2$, and $\sigma_\psi^2$ only affect the legitimate user but not the eavesdropper, the
corresponding HWIs are always detrimental to the ergodic secrecy rate. However, the additive distortion noise at the BS affects both the achievable ergodic rate of the MT and the capacity of the eavesdropper. Hence, it
is not a priori clear if this HWI is beneficial or detrimental to the ergodic secrecy rate. The following Proposition provides a criterion for judging the benefits of the additive BS distortion noise.

\textit{Proposition 4}: For time interval $t$, non-zero additive BS distortion noise with small $\kappa^{\rm BS}_t>0$,  $\kappa^{\rm BS}_t\to 0$, is beneficial for the achievable ergodic secrecy rate of the $k^{\rm th}$
MT if and only if
\begin{equation}\label{beneficial}
(1-\phi)[1-N_E/L-(1-N_E/L-N_E/K)\phi] \times \frac{1-N_E/L}{1-(1-2\phi)N_E/L}<\frac{\alpha \gamma (N \overline{\lambda}_k \phi+\gamma)}{\beta^2 \overline{\lambda}_k  N},
\end{equation}
where $\gamma = (a_k+c_k)\phi+\beta\mu_k(1-\phi)+\beta(\kappa^{\rm MT}_r+\xi^{\rm DL}/(\beta_k P_T))$.
\begin{IEEEproof}
For additive BS distortion noise to be beneficial for a given time interval $t$ and small $\kappa^{\rm BS}_t>0$, the derivative $\partial \underline{R}_k^{\rm sec}(t)/\partial \kappa^{\rm BS}_t$ at
$\kappa^{\rm BS}_t=0$ has to be positive. Assuming $\underline{R}_k^{\rm sec}(t)>0$, this condition leads to $\partial \underline{R}_k(t)/\partial \kappa^{\rm BS}_t|_{\kappa^{\rm BS}_t=0} >
\partial \overline{C}_E/\partial \kappa^{\rm BS}_t|_{\kappa^{\rm BS}_t=0}$, which can be further simplified to (\ref{beneficial}).
\end{IEEEproof}

\textit{Remark 2} : We note that the criterion in Proposition 4 only guarantees that additive BS distortion noise with small positive $\kappa^{\rm BS}_t$ is beneficial. The ergodic secrecy rate, $\underline{R}_k^{\rm sec}(t)$,
is in general not monotonic in $\kappa^{\rm BS}_t$ and larger $\kappa^{\rm BS}_t$ may be harmful even if small $\kappa^{\rm BS}_t$ are beneficial, see Section \ref{s5e}. Furthermore, since the right hand side of
(\ref{beneficial}) is always positive, we conclude that additive BS distortion noise with small $\kappa^{\rm BS}_t$ is always beneficial when $\phi=1$, i.e., when AN is not injected.

\section{Numerical Examples}\label{s5}
In this section, we provide numerical and simulation results to verify the analysis presented in Sections \ref{s3} and \ref{s4} and to illustrate the impact of HWIs on the ergodic secrecy rate. For the numerical results,
we numerically evaluate the analytical expression for the lower bound on the ergodic secrecy rate obtained by combining (\ref{rseck}), (\ref{Rkunder}), and (\ref{Cup1}). For the simulation results, we employ
Monte Carlo simulation and evaluate (\ref{rseck}) using $\underline{R}_k^{\rm sec}(t) = \log_2(1+\gamma_k(t))$ and $C_E{{(t)}}=\log_2(1+\gamma_E{{(t)}})$ with $\gamma_k(t)$ and $\gamma_E{{(t)}}$ given by (\ref{gammak})
and (\ref{gammaE}), respectively, for $5,000$ independent channel realizations. For simplicity, in this section, we assume that the path-loss for all MTs is identical \footnote{{Although the analytical results presented in this paper are valid for unequal path-losses, for the presented numerical results, we employ equal path-losses in order to be able to focus on the impact of HWIs on the achievable ergodic secrecy rate. The investigation of this impact is the main objective of this paper, and unequal path-losses do not provide any additional insights in this regard.}}, i.e., $\beta_k=1$, $1 \leq k \leq K$, and the coherence block length
is equal to $T=500$ time slots. Typical values for the phase noise increment standard deviations, $\sigma_\psi$, $\sigma_\phi$, used include $0.06^{\circ}$,  which was adopted in  the long-term evolution
(LTE) specifications \cite{LTE}, and $6^{\circ}$, which corresponds to strong phase noise according to \cite{strongpn1,strongpn2}. Furthermore, typical values for the
additive distortion noise $\kappa^{\rm MT}_t=\kappa^{\rm BS}_r=\kappa^{\rm BS}_t=\kappa^{\rm MT}_r$ include $\{0,0.05^2,0.15^2\}$ \cite{nonideal}, whereas the amplified receiver noise was set to $\xi^{\rm UL}=\xi^{\rm DL}=
1.58 \sigma^2_n$ \cite{MRC}, with $\sigma_n^2=1$. The specific values of the adopted system and HWI parameters  are provided in the captions of the figures.

\subsection{Capacity of Eavesdropper for G-NS AN Precoding} \label{s5a}
\begin{figure}[t]
 \centering
 \begin{minipage}[b]{0.45\linewidth} \hspace*{-1cm}
\includegraphics[width=3.7in]{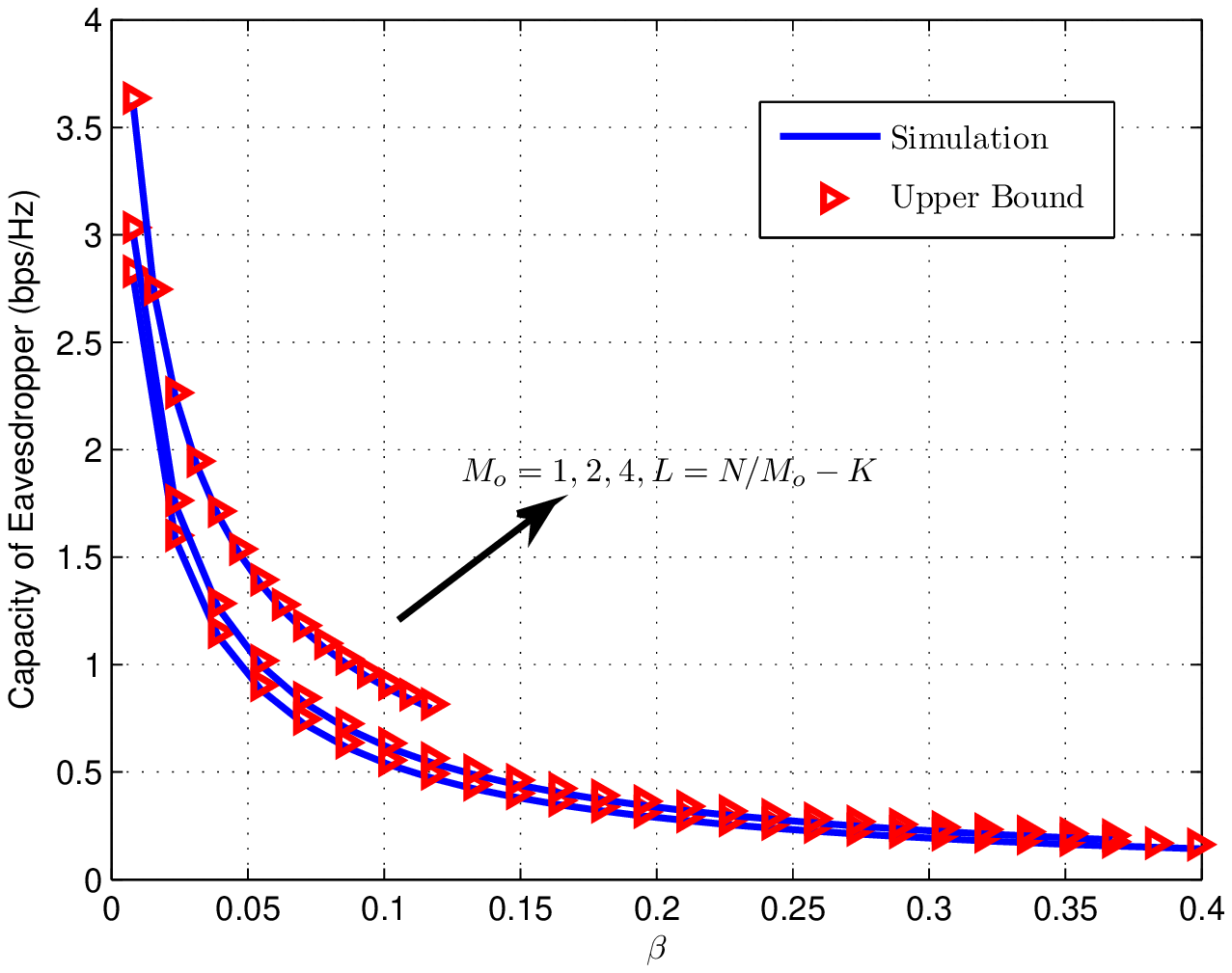}\vspace*{-8mm}\\[-6mm]
\caption{Capacity of the eavesdropper vs.~the normalized number of MTs $\beta$ for a system with $N=128$, $N_o=4$, $N_E=16$, $P_T=10$ dB, $\phi=0.25$, $\kappa^{\rm BS}_t=0.15^2$, and G-NS AN precoding
with $M_o=\{1,2,4\}$.}\label{fig1}
 \end{minipage}\hspace*{1.1cm}
 \begin{minipage}[b]{0.45\linewidth} \hspace*{-1cm}
\includegraphics[width=3.7in]{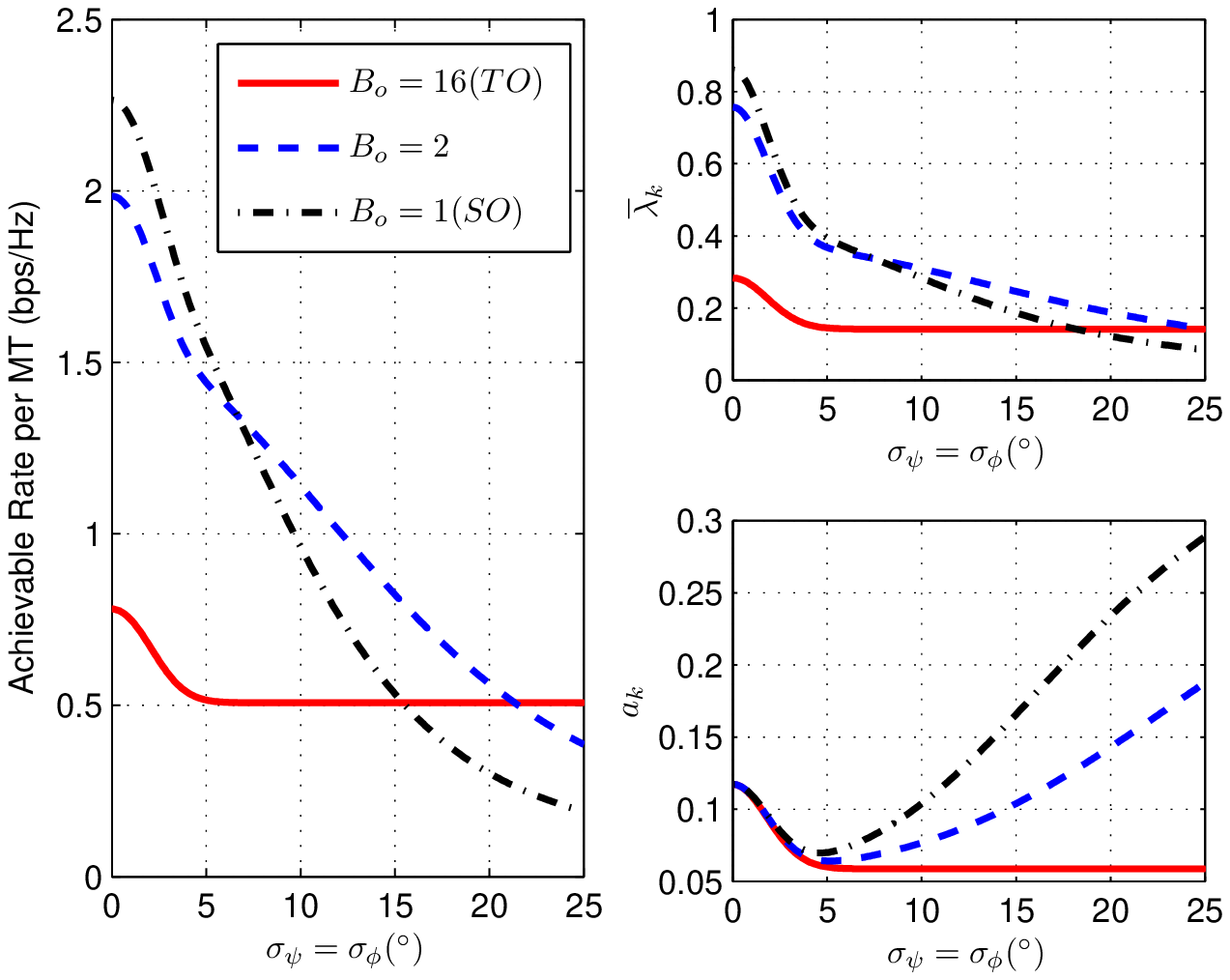}\vspace*{-8mm}\\[-6mm]
    \caption{Achievable ergodic rate, $\overline{\lambda}_k$, and $a_k$ vs.~phase noise standard deviation $\sigma_\psi=\sigma_\phi$ for different pilot designs for a system with $N=128$,
    $N_o=2$, $N_E=16$, $K=B=16$, $p_\tau=P_T/K$, $P_T=10$ dB, $\phi=0.5$, and $\kappa^{\rm BS}_t=\kappa^{\rm BS}_r=\kappa^{\rm MT}_t=\kappa^{\rm MT}_r=0.05^2$.}\label{fig12}
 \end{minipage}
\end{figure}
Fig.~\ref{fig1} depicts the eavesdropper's ergodic capacity, $C_E$, as a function of $\beta$ for G-NS AN precoding with $M_o=\{1,2,4\}$. Besides results for the analytical upper bound, $\overline{C}_E$,  from (\ref{Cup1}),
we also show simulation results for $C_E$ by averaging $\log_2(1+\gamma_E)$ over $5,000$ independent channel realizations, where $\gamma_E$ is given by (\ref{gammaE}). From Fig.~\ref{fig1} we observe that the proposed
upper bound on the capacity of the eavesdropper is very tight. Furthermore, as expected, the ergodic capacity of the eavesdropper is an increasing function of $M_o$ since the number of dimensions available for AN generation, $L
=N/M_o-K$, is a decreasing function of $M_o$. In fact, since $L=N/M_o-K>N_E$ is needed for successfully jamming the eavesdropper, for $M_o=4$, we depict the ergodic capacity of the eavesdropper only for $\beta<0.125$. Nevertheless, as will be shown below, choosing $M_o>1$ may still be beneficial as far as the ergodic secrecy rate is concerned as the achievable ergodic rate of the MT is an increasing function of $M_o$.

\subsection{Achievable Ergodic Rate of MT for Different Pilot Designs }\label{s5b}
Next, we investigate the impact of the general pilot designs introduced in Section \ref{s2a} on the lower bound of the achievable ergodic rate of the considered MT given in (\ref{Rkunder})\footnote{We note that all results
obtained by numerically evaluating the analytical expressions derived in this paper were verified by simulations. However, the simulation results are not included in all figures for clarity of presentation.}.
Note that the capacity of the eavesdropper is not affected by the pilot design. For simplicity, we assume equal duration for all training sub-phases,  $B_b=B/B_o$, $b\in\{1,\ldots,B_o\}$, and $B=K$.
The same number of users is assigned to each training sub-phase. In Fig.~\ref{fig12}, we show the achievable ergodic rate of a MT in training set ${\cal S}_{B_o}$ as well as the corresponding $\overline{\lambda}_k$,
which reflects the power of the received useful signal, and $a_k$, which reflects the multiuser interference. Results for $B_o=1$ (SO pilots), $B_o=2$, and $B_o=16$ (TO pilots) are shown. As predicted in Section
\ref{s4a}, the multiuser interference, $a_k$, is monotonically decreasing in $B_o$ as larger $B_o$ improve the robustness against phase noise during the channel estimation phase, which allows better suppression
of multiuser interference via MF precoding. Somewhat surprisingly, for $\sigma_\psi=\sigma_\phi \leq 5^\circ$, $a_k$ is a decreasing function of the phase noise variance. This may be attributed to the fact that phase noise
prevents the coherent superposition of the multiuser interference generated by different MTs such that large interference values are avoided. On the other hand, for $\sigma_\psi=\sigma_\phi > 5^\circ$, the detrimental
effects of the pilot {{interference}} caused by the loss of orthogonality for $B_o<16$ outweigh this positive effect and $a_k$ increases with the phase noise variance. For $\overline{\lambda}_k$, i.e., the received signal
power, we observe from Fig.~\ref{fig12} that the optimal $B_o$ depends on the phase noise variance. In particular, for small phase noise variances, small $B_o$ are preferable since the increased pilot power outweighs
the loss of orthogonality during training. On the other hand, for large phase noise variances, eventually TO pilots become optimal as the preserved orthogonality during training becomes
crucial. The behaviour of $\overline{\lambda}_k$ and $a_k$ is also reflected in the behaviour of the achievable rate of the considered MT. In particular, for the considered system parameters, $B_o=1$, $B_o=2$,
and $B_o=16$ are optimal for  $\sigma_\psi=\sigma_\phi \leq 6^\circ$, $6^\circ<\sigma_\psi=\sigma_\phi \leq 21^\circ$, and $\sigma_\psi=\sigma_\phi > 21^\circ$ (which is not a practical range), respectively.
Hence, in practice, the optimal $B_o$ can be found by evaluating (\ref{Rkunder}).

\subsection{Optimal Power Allocation to Data  and AN}\label{s5c}
\begin{figure}[t]
 \begin{minipage}[b]{0.45\linewidth} \hspace*{-1cm}
\includegraphics[width=3.7in]{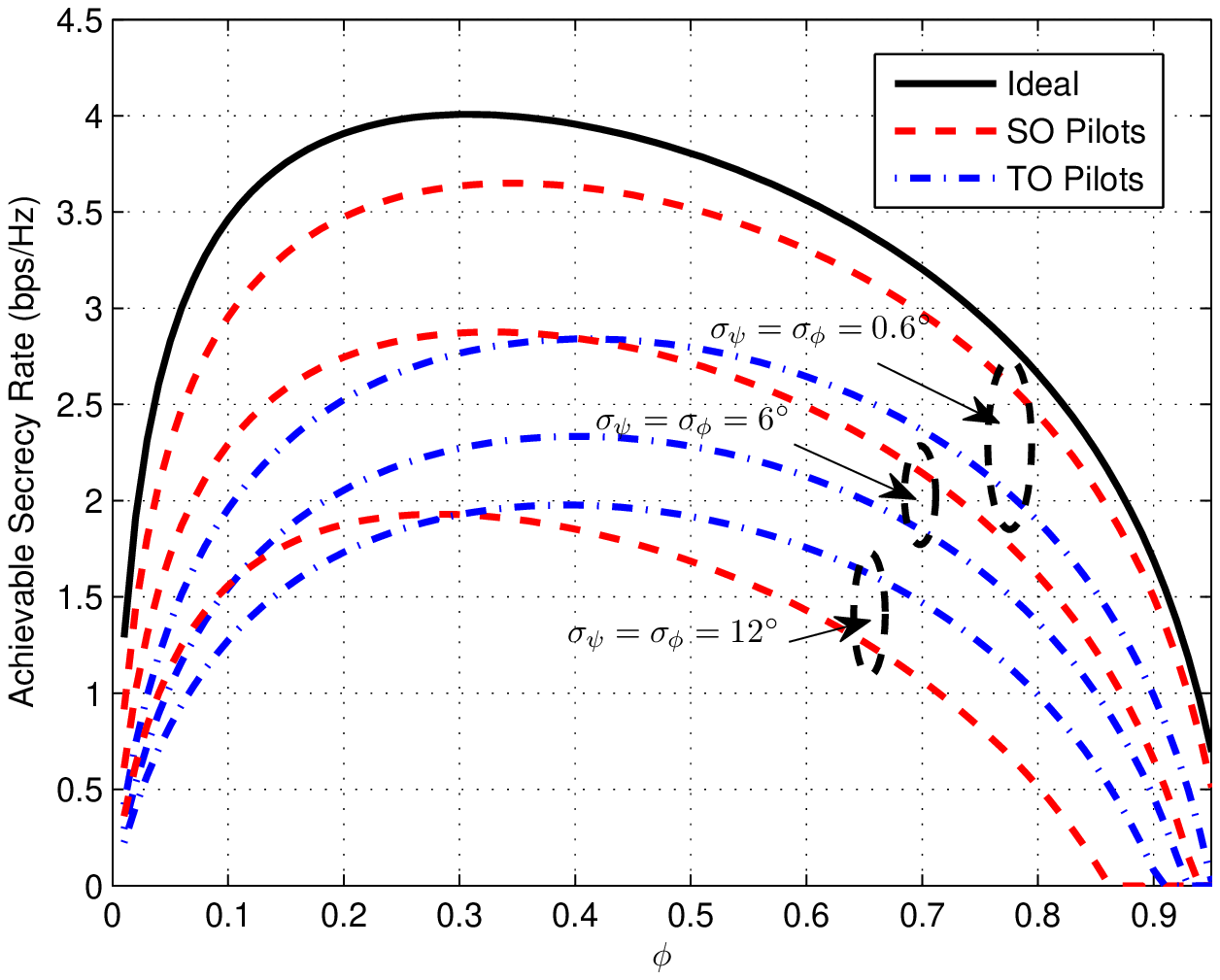}\vspace*{-8mm}\\[-6mm]
    \caption{Achievable ergodic secrecy rate vs.~$\phi$ for SO and TO pilots and  a system with $K=B=4$ , $N=128$, $N_o=M_o=2$, $N_E=4$, $p_\tau=P_T/K$, $P_T=10$ dB, and $\kappa^{\rm BS}_t=\kappa^{\rm BS}_r=\kappa^{\rm MT}_t=\kappa^{\rm MT}_r=0.15^2$.}\label{fig4}
 \end{minipage}\hspace*{1.1cm}
 \begin{minipage}[b]{0.45\linewidth} \hspace*{-1cm}
\includegraphics[width=3.7in]{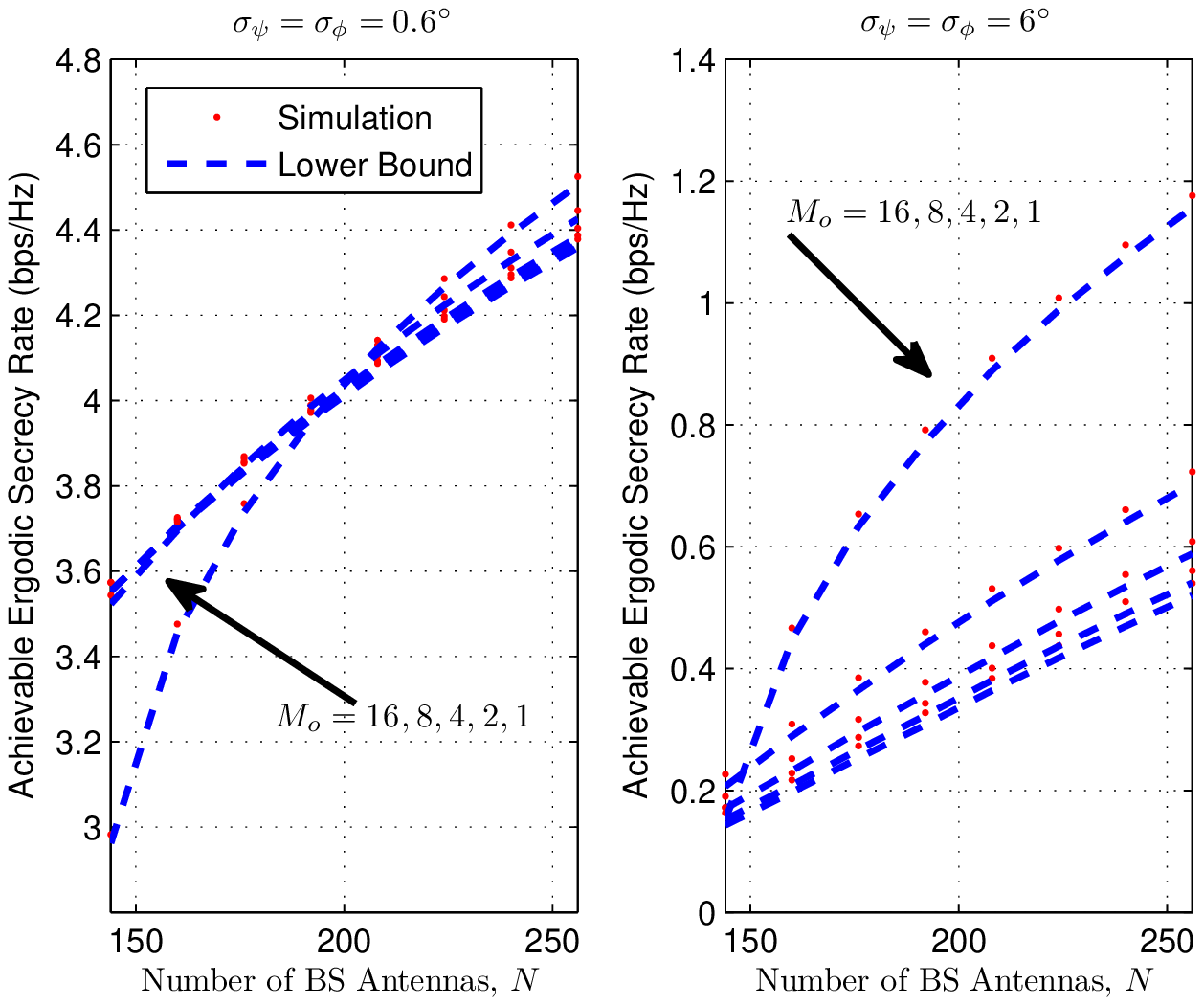}\vspace*{-8mm}\\[-6mm]
    \caption{Achievable ergodic secrecy rate vs.~number of BS antennas for G-NS AN precoding and a system with $K=B=4$, $N_E=4$, $N_o=16$, $B_o=1$, $p_\tau=P_T/K$, $P_T=10$ dB, and $\kappa^{\rm BS}_t=\kappa^{\rm BS}_r=\kappa^{\rm MT}_t=\kappa^{\rm MT}_r=0.15^2$. The optimal $\phi$ is adopted.}\label{fig3}
     \end{minipage}\hspace*{1.1cm}
\end{figure}

Fig.~\ref{fig4} shows the achievable ergodic secrecy rate as a function of the power allocation parameter $\phi$ for SO and TO pilots and different phase noise variances. G-NS AN precoding with $M_o=N_o=2$ is adopted. The curve for ideal hardware components, i.e., $\kappa^{\rm BS}_t=\kappa^{\rm BS}_r=\kappa^{\rm MT}_t=\kappa^{\rm MT}_r=\sigma_\psi=\sigma_\phi=0$, is also provided for reference. We investigate the optimal power allocation between data transmission and AN emission for the maximization of the ergodic secrecy rate achieved for different phase noise levels. When the phase noise variance is small, i.e., $\sigma_\psi=\sigma_\phi=0.6^\circ$, SO pilots outperforms TO pilots for all values of $\phi$. However, this is not true for stronger phase noise. We also observe that the optimal value for $\phi$ maximizing the ergodic secrecy rate is only weakly dependent on the
phase noise variance.

\subsection{Achievable Ergodic Secrecy Rate with G-NS AN Precoding}\label{s5d}
In Fig. \ref{fig3}, we show the ergodic secrecy rate achieved with G-NS AN precoding for different values of $M_o$ as a function of the number of BS antennas. The cases of weak ($\sigma_\psi=\sigma_\phi=0.6^\circ$)
and strong ($\sigma_\psi=\sigma_\phi=6^\circ$) phase noise are considered. For weak phase noise, using large values of $M_o$ becomes beneficial only for large numbers of antennas, i.e., $N>200$, as for smaller numbers
of antennas the positive effect of larger values of $M_o$ on the AN leakage is outweighed by their negative effect on the number of spatial dimensions available for AN precoding. On the other hand, for strong phase noise,
the AN leakage is larger and its mitigation by choosing $M_o=N_o=16$ is beneficial already for $N>150$. These observations are in line with our theoretical considerations in Section \ref{s4b}. Fig.~\ref{fig3} also
confirms the accuracy of the derived analytical expressions for the ergodic secrecy rate.

\subsection{Maximum Tolerable Number of Eavesdropper Antennas}\label{s5e}
\begin{figure}[t]
 \begin{minipage}[b]{0.45\linewidth} \hspace*{-1cm}
\includegraphics[width=3.7in]{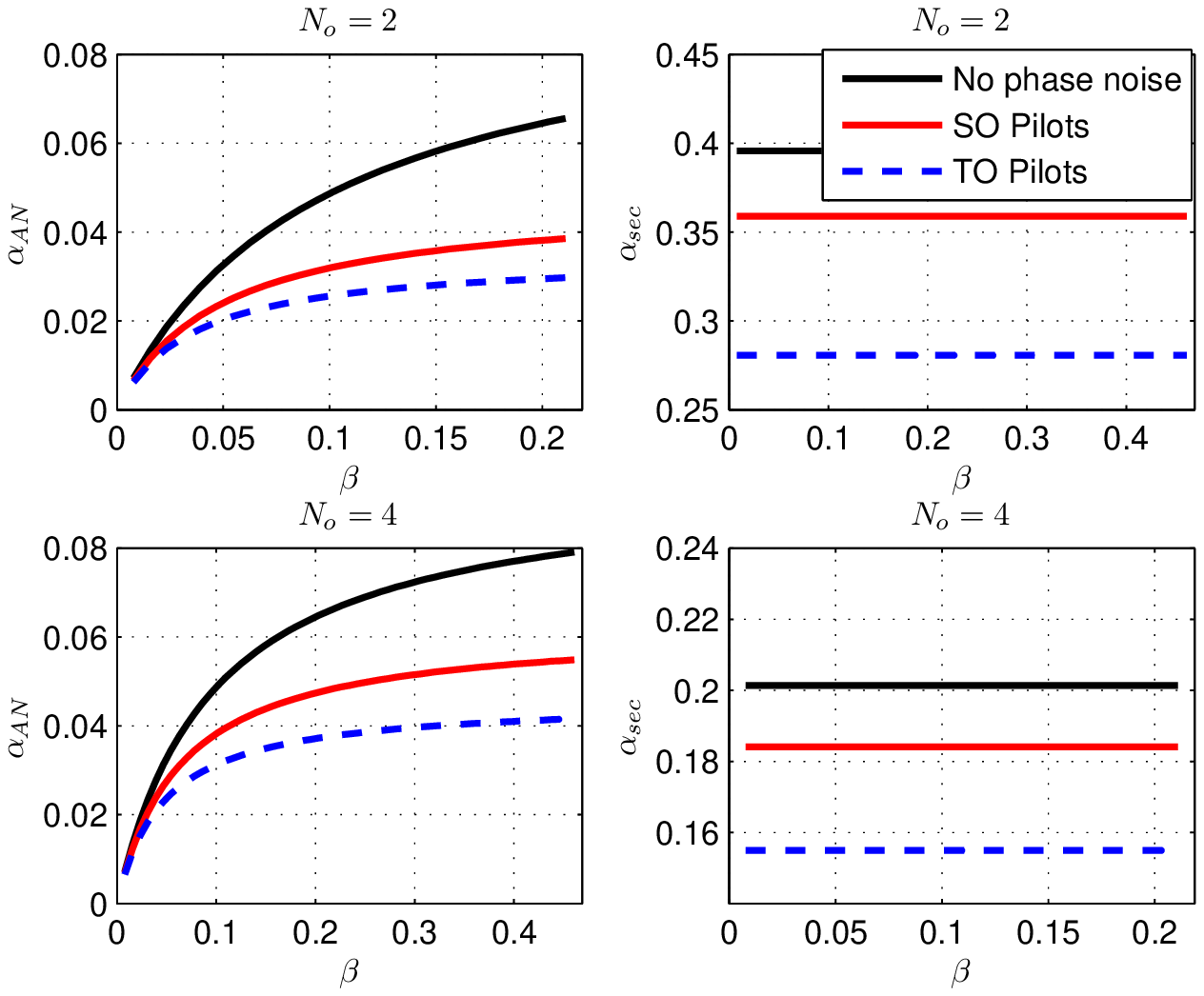}\vspace*{-8mm}\\[-6mm]
    \caption{$\alpha_{\rm AN}$ and $\alpha_{\rm sec}$ vs. the normalized number of MTs $\beta$ for SO and TO pilots and a system with $N=128$, $M_o=2$,
    $p_\tau=P_T/K$, $P_T=10$ dB, $\sigma_\psi=\sigma_\phi=6^\circ$, and $\kappa^{\rm BS}_t=\kappa^{\rm BS}_r=\kappa^{\rm MT}_t=\kappa^{\rm MT}_r=0.15^2$.}\label{fig5}
 \end{minipage}\hspace*{1.1cm}
\begin{minipage}[b]{0.45\linewidth} \hspace*{-1cm}
\includegraphics[width=3.7in]{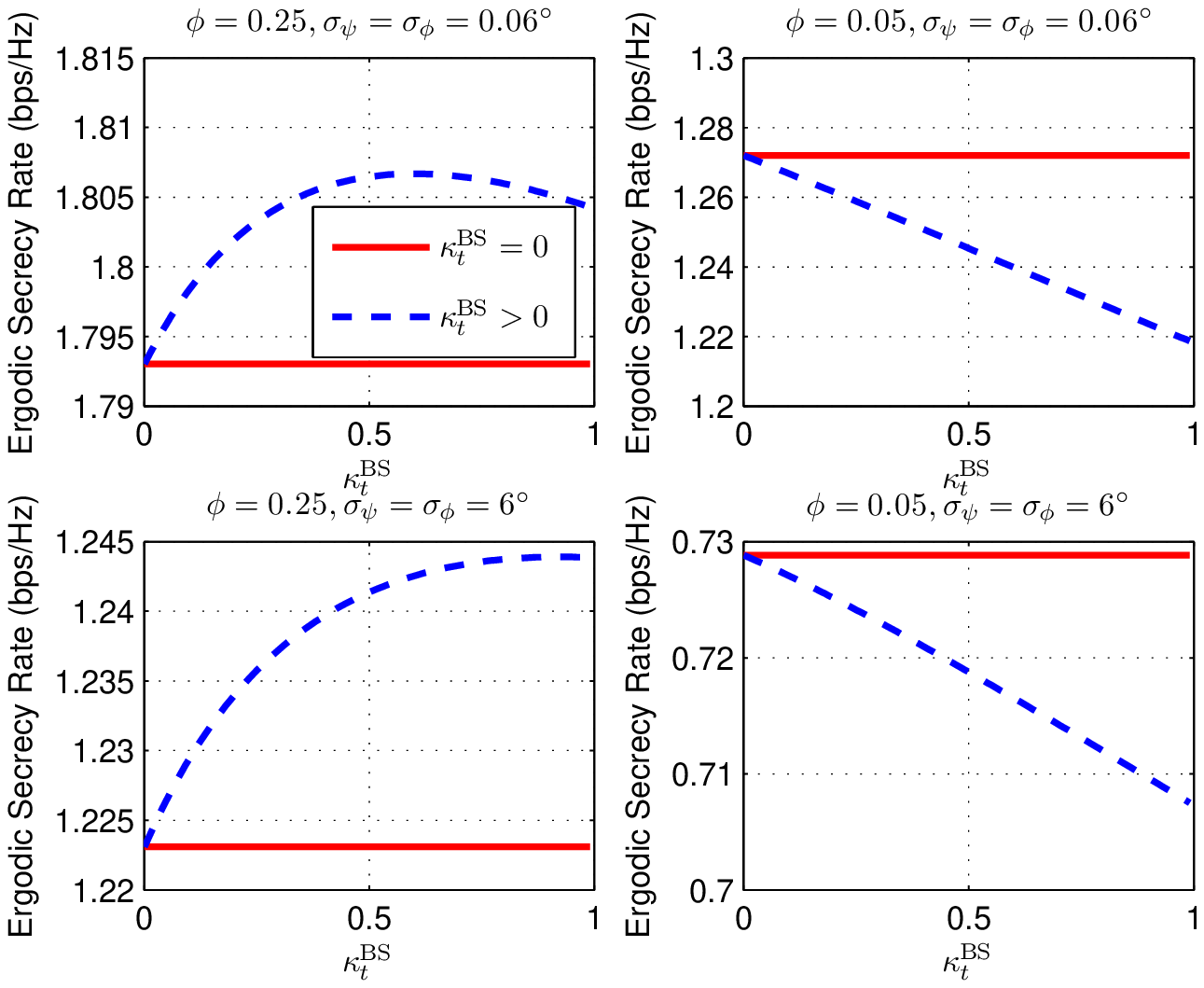}\vspace*{-8mm}\\[-6mm]
\caption{Achievable ergodic secrecy rate vs.~BS distortion noise parameter $\kappa^{\rm BS}_t$ for a system with $N=128$, $K=B=32$, $N_E=4$, $N_o=M_o=2$, $p_\tau=P_T/K$, $P_T=10$ dB, and
$\kappa^{\rm BS}_r=\kappa^{\rm MT}_t=\kappa^{\rm MT}_r=0.15^2$.}\label{fig6}
 \end{minipage}\hspace*{1.1cm}
\end{figure}
Fig.~\ref{fig5} depicts the (normalized) maximum tolerable number of eavesdropper antennas for achieving a positive ergodic secrecy rate for the case without AN generation, $\alpha_{\rm AN}$, and the case with AN
generation,  $\alpha_{\rm sec}$, as a function of the (normalized) number of users, $\beta$. Results for channel estimation based on SO and TO pilots as well as the case of no phase noise ($\sigma_\psi=\sigma_\phi=0^\circ$) are
shown for $N_o=2$ and $N_o=4$ LOs. First, we note that, as expected from our considerations in Section \ref{s4f},  for the case without AN ($\phi=1$), increasing $N_o$ from 2 to 4 is
beneficial, i.e., the number of tolerable eavesdropper antennas increases. In contrast, if AN is injected, $N_o=2$ is preferable. Second, AN generation is beneficial and improves the robustness against eavesdropping,
i.e., $\alpha_{\rm sec}>\alpha_{\rm AN}$. Third, as expected from Sections \ref{s4c} and  \ref{s4d},  $\alpha_{\rm AN}$ is a monotonically increasing function of $\beta$ whereas $\alpha_{\rm sec}$ is independent of
$\beta$. Fourth, for the considered example of weak phase noise, SO pilots outperform the TO pilots for all considered cases.

\subsection{Is Additive Distortion Noise at the BS Beneficial for Security?}\label{s5f}
In Fig.~\ref{fig6}, we show the achievable ergodic secrecy rate as a function of the BS distortion noise parameter, $\kappa^{\rm BS}_t$, for different phase noise variances and different power allocation factors $\phi$.
For comparison, the achievable ergodic secrecy rates without BS distortion noise (i.e., $\kappa^{\rm BS}_t=0$) are also shown. Fig.~\ref{fig6} shows that if the power allocated to AN is substantial (e.g., $\phi=0.05$), the additional distortion noise has a negative effect on the ergodic secrecy rate. On the other hand, if the power assigned for AN is not sufficient (e.g., $\phi=0.25$), non-zero additive distortion noise at the BS is beneficial as the distortion
noise acts like additional AN. In particular, for $\phi=0.25$, $\sigma_\psi=0.06^\circ$, we obtain for the left hand side and right hand side of (\ref{beneficial}) $0.52$ and $1.66$, respectively, which we represent as $(0.52, 1.66)$.
Correspondingly, we obtain for $\phi=0.25$, $\sigma_\psi=6^\circ$ and $\phi=0.05$, $\sigma_\psi=0.06^\circ$ and $\phi=0.05$, $\sigma_\psi=6^\circ$ the tupels $(0.52, 2.53)$ and $(0.80, 0.16)$ and $(0.80, 0.35)$, respectively.
These values and the results in Fig.~\ref{fig6} suggest that  (\ref{beneficial}) can indeed be used to predict whether or not BS distortion noise is beneficial.

\section{Conclusions}\label{s6}
In this paper, we have investigated the impact of HWIs such as multiplicative phase noise, additive distortion noise, and amplified receiver noise on the secrecy performance of massive MIMO systems employing
MF precoding for downlink data transmission. To mitigate the loss of pilot orthogonality during uplink training if multiple MTs emit pilots concurrently, a generalized pilot design was proposed. Furthermore, to avoid the
AN leakage caused by the loss of orthogonality between the user channels and the NS AN precoder if multiple noisy LOs are employed at the BS, a novel G-NS AN precoding scheme was introduced. For the considered
system, a lower bound on the achievable ergodic secrecy rate of the users was derived. This bound was used to obtain important insights for system design, including the impact of the pilot sequence design, the AN
precoder design, the number
of LOs, and the various HWI parameters. The following general conclusions can be drawn: 1) Additive distortion noise at the BS may be beneficial for the secrecy performance especially if little or no AN is injected; 2)
all other HWIs have a negative impact on the ergodic secrecy rate; 3) despite their susceptibility to pilot {{interference}} in the presence of phase noise, SO pilots are preferable except for the case when
the phase noise is very strong; 4) if the number of BS antennas is sufficiently large, the proposed G-NS AN precoder outperforms the conventional NS AN precoder in the presence of phase noise.

{Interesting extensions of this paper which could be studied in future research include the impact of HWIs on the physical layer security of multi-cell massive MIMO systems, pilot sequence design under an average power constraint, and optimal AN precoder design for secrecy rate maximization under HWIs.}
\section*{Appendix}
\subsection{A Useful Theorem from Free Probability Theory}
\textit{Theorem 1}\cite{free}: If $({\bf U},{\bf V})\in \mathbb{C}^{N}$ are free from $({\bf Y},{\bf Z})\in \mathbb{C}^{N}$, then ${\rm Tr}\left({\bf UYVZ}\right)=$
\begin{equation}
{\rm Tr}\left({\bf U}\right){\rm Tr}\left({\bf V}\right){\rm Tr}\left({\bf YZ}\right)+{\rm Tr}\left({\bf Y}\right){\rm Tr}\left({\bf Z}\right){\rm Tr}\left({\bf UV}\right)-{\rm Tr}\left({\bf U}\right){\rm Tr}\left({\bf V}\right){\rm Tr}\left({\bf Y}\right){\rm Tr}\left({\bf Z}\right),
\end{equation}
where ${\rm Tr}\left(\cdot\right)=\lim_{N \to \infty} {\rm tr}\left(\cdot\right)/N$.

\subsection{Proof of {Lemma 1}}
The ergodic secrecy rate achieved by the $k^{\rm th}$ MT in symbol interval $t \in \{B+1,\ldots,T\}$ is given by \cite[{Lemma 1}]{zhu}
\begin{equation}
R^{\rm sec}_k(t)=\mathbb{E}\left[[R_k(t)-\log_2(1+\gamma_E{{(t)}})]^+\right] {\geq} \left[\mathbb{E}[R_k(t)]-C_E{{(t)}}\right]^+ \overset{(a)}{\geq} \left[\underline{R}_k(t)-C_E{{(t)}}\right]^+
=\underline{R}^{\rm sec}_k(t),
\end{equation}
where $\underline{R}^{\rm sec}_k(t)$ is an achievable lower bound for $R^{\rm sec}_k(t)$, and $(a)$ uses (\ref{Rkt}). By averaging $R^{\rm sec}_k(t)$ over
all symbol intervals $t \in \{B+1,\ldots,T\}$ we obtain Lemma 1. This completes the proof.

\subsection{Proof of {Lemma 2}}
The expectation given in (\ref{lambdak}) for $k \in {\cal S}_b$ is calculated as $\mathbb{E}\bigg[{\bf g}^H_k \boldsymbol{\Theta}^H_k(t) {\bf f}_k\bigg]$
\begin{eqnarray}\label{sig1}
\nonumber &\overset{(a)}{=}& \mathbb{E}\bigg[\frac{\hat{\bf g}^H_k \boldsymbol{\Psi}^H_{t_0}(t) \hat{\bf g}_{k}}{\|\hat{\bf g}_{k}\|}e^{j(\phi_k(t)-\phi_k(t_0))}\bigg]
\overset{(b)}{=} {\rm tr}\left(\mathbb{E}\bigg[\frac{\hat{\bf g}_{k}\hat{\bf g}^H_k}{\|\hat{\bf g}_{k}\|} \bigg] \mathbb{E}\bigg[\boldsymbol{\Psi}^H_{t_0}(t)\bigg]\right) \mathbb{E}\bigg[e^{j(\phi_k(t)-\phi_k(t_0))}\bigg]\\
&{=}&\sqrt{\beta_k N \lambda_k} \cdot e^{-\frac{\sigma^2_\psi+\sigma^2_\phi}{2}|t-t_0|},
\end{eqnarray}
where $\boldsymbol{\Psi}_{t_0}(t)={\rm diag}\left(e^{j (\psi_1(t)-\psi_1(t_0))} {\bf 1}^T_{1 \times N/N_o},\ldots,e^{j (\psi_{N_o}(t)-\psi_{N_o}(t_0))}{\bf 1}^T_{1 \times N/N_o}\right)$ and $\lambda_k$ is defined in {Lemma 2}. In (\ref{sig1}), $(a)$ exploits that the channel estimate and the estimation error are uncorrelated \cite{nonideal}, and $(b)$ exploits the mutually independence of $\hat{\bf g}_{k}\hat{\bf g}^H_k$,
$\boldsymbol{\Psi}^H_{t_0}(t)$, and $e^{j(\phi_k(t)-\phi_k(t_0))}$. This completes the proof.

\subsection{Proof of {Lemma 3}}
In (\ref{gammak}), the term reflecting the interference caused by the signal intended for MT $l \in {\cal S}_b$ to MT $k \in {\cal S}_b$ can be expanded as
$\mathbb{E}\left[\left|{\bf g}^H_k \boldsymbol{\Theta}^H_k(t) {\bf f}_l\right|^2\right]=$
\begin{eqnarray}\label{int}
\nonumber && \mathbb{E}\left[\left|{\bf g}^H_{k}(t_0) \boldsymbol{\Psi}^H_{t_0}(t) \frac{\hat{\bf g}_{l}}{\|\hat{\bf g}_{l}\|}e^{j(\phi_k(t)-\phi_k(t_0))}\right|^2\right]
=\mathbb{E}\left[{\rm tr}\left({\bf g}_{k}(t_0){\bf g}^H_{k}(t_0)\boldsymbol{\Psi}^H_{t_0}(t)\frac{\hat{\bf g}_{l}\hat{\bf g}^H_l}{\|\hat{\bf g}_{l}\|^2}\boldsymbol{\Psi}_{t_0}(t)\right)\right]\\
&\overset{(a)}{=}& \beta_k+\left(\frac{\mathbb{E}\bigg[{\rm tr}\left({\bf X}^H_l {\bf g}_k(t_0) {\bf g}^H_k(t_0) {\bf X}_l \boldsymbol{\psi}_b \boldsymbol{\psi}_b^H\right)\bigg]}{\beta_l^2 \boldsymbol{\omega}_l^H \boldsymbol{\Theta}^b_{\sigma(t_0)} \boldsymbol{\Sigma}_b^{-1} \boldsymbol{\Theta}^H_{\sigma(t_0)} \boldsymbol{\omega}_l N} -\beta_k\right) \mathbb{E}_\psi\bigg[\left(\frac{1}{N}{\rm tr}\left( \boldsymbol{\Psi}^H_{t_0}(t) \right)\right)^2\bigg],
\end{eqnarray}
where ${\bf X}_l=\beta_l \boldsymbol{\omega}^H_l \boldsymbol{\Theta}^b_{\sigma(t_0)} \boldsymbol{\Sigma}_b^{-1} \otimes {\bf I}_N$, and $(a)$ exploits Theorem 1 from free probability theory, since the phase drift matrices $\boldsymbol{\Psi}_{t_0}(t)$ and $\boldsymbol{\Psi}^H_{t_0}(t)$ are free from ${\bf g}_{k}(t_0) {\bf g}^H_{k}(t_0)$ and $\frac{\hat{\bf g}_{l}\hat{\bf g}^H_l}{\|\hat{\bf g}_{l}\|^2}$. For notational simplicity,
we define $I =\mathbb{E}\left[{\rm tr}\left({\bf X}^H_l {\bf g}_k(t_0) {\bf g}^H_k(t_0) {\bf X}_l \boldsymbol{\psi}_b \boldsymbol{\psi}_b^H\right)\right]$, which can be further expanded as
\begin{eqnarray}\label{I}
\nonumber I&=&\mathbb{E}\left[{\rm tr}\left({\bf Y}^H_{lk} {\bf g}_k {\bf g}^H_k {\bf Y}_{lk} {\bf g}_k {\bf g}^H_k\right)\right]+ {\rm tr}\left(\beta_k {\bf X}^H_l{\bf X}_l (\boldsymbol{\Sigma}_b-\beta_k\left({\bf W}_k^b+{\bf U}^b_k\right)\right) \otimes {\bf I}_N)+ \\
&& \mathbb{E}\left[{\rm tr}\left({\bf X}^H_l {\bf g}_k {\bf g}^H_k {\bf X}_l \left({\bf U}^b_k \otimes {\rm diag} \left(g^{(1)}_k,\ldots,g^{(N)}_k\right)\right)\right)\right],
\end{eqnarray}
where
\begin{equation}
{\bf Y}_{lk}=\boldsymbol{\Theta}^H_k(t_0) {\bf X}_l \left[\boldsymbol{\Theta}^H_k(\overline{B}_{b-1}+1) \omega_k(\overline{B}_{b-1}+1),\ldots,\boldsymbol{\Theta}^H_k(t_0) \omega_k(t_0) \right]^T.
\end{equation}
Denoting the $t^{\rm th}$ column of ${\bf I}_N$ by ${\bf e}_t^N \in \mathbb{C}^{N \times 1}$, the first term on the right hand side of (\ref{I}), denoted by $I_1$, can be expanded as
\begin{eqnarray}\label{I1}
\nonumber I_1 &=& \sum_{n_1,n_2,b_1,b_2} [\beta_k {\bf X}_l{\bf e}_{b_1}^{B_b} \otimes {\bf I}_N]_{n_1 n_1} [\beta_k {\bf X}_l{\bf e}_{b_2}^{B_b} \otimes {\bf I}_N]^H_{n_2 n_2} \times \omega_k({b_1}) \omega^*_k({b_2}) \Theta(n_1,n_2,b_1,b_2,t_0)\\
\nonumber &=&\left|{\rm tr}\left(\beta_k {\bf X}_l (\boldsymbol{\Theta}^b_{\sigma(t_0)} \boldsymbol{\omega}_k \otimes {\bf I}_N)\right)\right|^2 + {\rm tr}\left(\beta^2_k {\bf X}^H_l {\bf X}_l ({\bf W}_k^b \otimes {\bf I}_N)\right)\\
&+& \sum^N_{|n_1-n_2| \leq \frac{N}{N_0}}\beta_k^2 ({\bf e}_{n_1}^N)^H {\bf X}_l \left(({\bf W}_k^b-\boldsymbol{\Theta}^b_{\sigma(t_0)} \boldsymbol{\omega}_k \boldsymbol{\omega}_k^H \boldsymbol{\Theta}^b_{\sigma(t_0)})\otimes {\bf e}_{n_1}^N({\bf e}^N_{n_2})^H\right) {\bf X}_l^H {\bf e}_{n_2}^N,
\end{eqnarray}
where the expectation with respect to the phase drift, $\Theta(n_1,n_2,b_1,b_2,t_0)$, depends on the number of LOs, $N_o$, and is given by $\Theta(n_1,n_2,b_1,b_2,t_0)=$
\begin{equation}
\mathbb{E}\bigg[e^{\theta_k^{n_1}({b_1})-\theta_k^{n_1}(t_0)-\theta_k^{n_2}(b_2)+\theta_k^{n_2}(t_0)}\bigg]=\begin{cases} e^{-\frac{\sigma^2_\psi+\sigma^2_\phi}{2}|{b_1}-{b_2}|} & |n_1-n_2| \leq \frac{N}{N_o},\\
e^{-\frac{\sigma^2_\psi+\sigma^2_\phi}{2}|t_0-{b_1}|} e^{-\frac{\sigma^2_\psi+\sigma^2_\phi}{2}|t_0-{b_2}|} & |n_1-n_2| > \frac{N}{N_o}.\end{cases}
\end{equation}
Furthermore, we rewrite ${\bf U}^b_k=(\kappa_t^{\rm MT}+\kappa_r^{\rm BS}) p_\tau \sum_{t=1}^{B_b} {\bf e}_t^{B_b} ({\bf e}_t^{B_b})^H$ and ${\rm diag} \left(g^{(1)}_k,\ldots,g^{(N)}_k\right)=\sum_{n=1}^N
|({\bf e}^N_n)^H {\bf g}_k|^2 {\bf e}_n^N ({\bf e}^N_n)^H$. Using these results in the third term on the right hand side of (\ref{I}), denoted by $I_2$, we obtain
\begin{equation}\label{I3}
I_2=\beta_k^2{\rm tr}\left({\bf X}^H_l {\bf X}_l ({\bf U}^b_k \otimes {\bf I}_N)\right)+\sum_{n=1}^N \beta_k^2 ({\bf e}^N_n)^H {\bf X}_l \left({\bf U}^b_k \otimes {\bf e}_n^N ({\bf e}^N_n)^H \right) {\bf X}_l {\bf e}_n^N.
\end{equation}
Applying (\ref{I1}) and (\ref{I3}) in (\ref{int}) and exploiting $\mathbb{E}\bigg[\left(\frac{1}{N}{\rm tr}\left( \boldsymbol{\Psi}^H_{t_0}(t) \right)\right)^2\bigg]=\frac{1-\epsilon}{N_o}+\epsilon$, we obtain the result in Lemma
3 for $k,l \in {\cal S}_b$.

For the case of $l \notin {\cal S}_b$, the multiuser interference term simplifies to
\begin{equation}\label{int2}
\mathbb{E}\left[\left|{\bf g}^H_k \boldsymbol{\Theta}^H_k(t) {\bf f}_l\right|^2\right]=\mathbb{E}\left[\left|{\bf g}^H_{k}(t_0)  \boldsymbol{\Psi}^H_{t_0}(t) \frac{\hat{\bf g}_{l}}{\|\hat{\bf g}_{l}\|}e^{j(\phi_k(t)-\phi_k(t_0) )}\right|^2\right]=\beta_k,
\end{equation}
where the last equality follows from the independence of ${\bf g}_{k}$, $\hat{\bf g}_{l}$, $l \notin {\cal S}_b$, and $\boldsymbol{\Psi}^H_{t_0}(t)$. This completes the proof.

\subsection{Proof of Lemma 4}\label{proofl4}
The AN leakage power received at the $k^{\rm th}$ MT in time slot $t$ can be expanded as
\begin{equation}\label{l1}
L^k_{\rm AN}(t)=\mathbb{E}\left[{\rm tr}\left(\hat{\bf g}_{k}\hat{\bf g}^H_k\boldsymbol{\Psi}^H_{t_0}(t){\bf AA}^H \boldsymbol{\Psi}_{t_0}(t)\right)\right]+
\mathbb{E}\left[{\bf e}_k^H(t_0) \boldsymbol{\Psi}^H_{t_0}(t) {\bf A}{\bf A}^H \boldsymbol{\Psi}^H_{t_0}(t) {\bf e}_k(t_0)\right].
\end{equation}
By using Theorem 1, the first term in (\ref{l1}) can be further expanded as
\begin{equation}\label{l2}
\beta_k L+\left(\mathbb{E}\left[{\rm tr}\left(\hat{\bf g}_{k} \hat{\bf g}^H_k {\bf AA}^H \right)\right] -\beta_k L\right) \mathbb{E}_\psi\left[\left(\frac{1}{N}{\rm tr}\left( \boldsymbol{\Psi}_{t_0}(t) \right)\right)^2\right]
=\beta_k L\left(1-\frac{1}{N_o}\right)\left(1-\epsilon\right),
\end{equation}
since phase drift matrices $\boldsymbol{\Psi}_{t_0}(t)$ and $\boldsymbol{\Psi}^H_{t_0}(t)$ are free from $\hat{\bf g}_{k} \hat{\bf g}^H_k$ and ${\bf AA}^H$. Furthermore, we exploited
$\hat{\bf g}^H_k {\bf A}={\bf 0}$, $1 \leq k \leq K$, which holds for the NS AN precoder.

The second term in (\ref{l1}) is equal to $\beta_k L (1-\lambda_k)$, with $\lambda_k$ as defined in Lemma 2, due to the mutual independence of the estimation error vector ${\bf e}_k(t_0)$, the phase drift matrix
$\boldsymbol{\Psi}_{t_0}(t)$, and the AN precoder ${\bf A}$. Combining these two terms completes the proof.

\subsection{Proof of Lemma 5}
For the G-NS AN precoder, we rewrite the leakage power received at the $k^{\rm th}$ MT in time slot $t$ as
\begin{equation}\label{l3}
L^k_{\rm AN}=\sum_{m=1}^{M_o}\mathbb{E}\left[\left({\bf g}^{(m)}_k\right)^H \left(\boldsymbol{\Theta}^{(m)}_k(t)\right)^H {\bf A}_{(m)}{\bf A}^H_{(m)} \boldsymbol{\Theta}^{(m)}_k(t) {\bf g}^{(m)}_k\right],
\end{equation}
where ${\bf g}^{(m)}_{k} \in \mathbb{C}^{N/M_o \times 1}$ contains the $((m-1)N/M_o+1)^{\rm th}$ to the $(m N/M_o)^{\rm th}$ elements of vector ${\bf g}_{k}$, $1 \leq m \leq M_o$, and $\boldsymbol{\Theta}^{(m)}_k(t)
\in \mathbb{C}^{N/M_o \times N/M_o}$ is a diagonal matrix with the $((m-1)N/M_o+1)^{\rm th}$ to the $(m N/M_o)^{\rm th}$ elements of matrix $\boldsymbol{\Theta}_k(t)$ on its main diagonal. Using similar steps as in
Appendix E but with $N_o/M_o$ substituted by $N_o$ for calculation of the expectation terms in (\ref{l3}), we obtain (\ref{AN2}). This completes the proof.

\subsection{Proof of Proposition 1}
{{
We first adopt Jensen's inequality to upper bound the eavesdropper's capacity at time interval $t$, as $C_E(t) \leq \log_2 \left(1+\mathbb{E}[\gamma_E(t)]\right)$, with $\gamma_E(t)$ given in (15). One further step to simplify $\mathbb{E}[\gamma_E(t)]$ requires the statistical independence between ${\bf g}^k_E(t)$ and the matrix ${\bf X}={\bf G}_E^H \boldsymbol{\Psi}^H(t)(q{\bf A}{\bf A}^H+\boldsymbol{\Upsilon}^{\rm BS}_t) \boldsymbol{\Psi}(t){\bf G}_E$. We note that the entries of ${\bf G}_E$ are independent complex Gaussian random variables. On the other hand, for the G-NS AN precoder, the columns of $\boldsymbol{\Psi}^H(t){\bf A} \in \mathbb{C}^{N \times L}$ form an orthonormal basis. Hence, ${\bf G}_E^H \boldsymbol{\Psi}^H(t){\bf A}$ also has independent complex Gaussian entries, which are independent from the entries of ${\bf g}^k_E(t)$ [9]. Besides, the term ${\bf G}^H_E  \boldsymbol{\Psi}^H(t) \boldsymbol{\Upsilon}^{\rm BS}_t \boldsymbol{\Psi}(t) {\bf G}_E$ converges to a deterministic diagonal matrix for $N \to \infty$, which is obviously independent of ${\bf g}^k_E(t)$. Therefore, $\mathbb{E}\bigg[{\bf g}^k_E(t) {\bf X}^{-1} ({\bf g}^k_E(t))^H\bigg]$ can be rewritten as $\mathbb{E}\bigg[{\bf g}^k_E(t) \mathbb{E}[{\bf X}^{-1}] ({\bf g}^k_E(t))^H\bigg]$, with $\mathbb{E}[{\bf X}^{-1}]$ approximated as a scaled identity matrix as in \cite[Appendix C]{zhu}. This leads to the upper bound given in (\ref{Cup1}).}}

\end{document}